# Determining Possible and Necessary Winners under Common Voting Rules Given Partial Orders

**Lirong Xia**                                                                                   lxia@cs.duke.edu
**Vincent Conitzer**                                                                          conitzer@cs.duke.edu
*Department of Computer Science, Duke University,*
*Durham, NC 27708, USA*

## Abstract

Usually a voting rule requires agents to give their preferences as linear orders. However, in some cases it is impractical for an agent to give a linear order over all the alternatives. It has been suggested to let agents submit partial orders instead. Then, given a voting rule, a profile of partial orders, and an alternative (candidate) $c$, two important questions arise: first, is it still possible for $c$ to win, and second, is $c$ guaranteed to win? These are the *possible winner* and *necessary winner* problems, respectively. Each of these two problems is further divided into two sub-problems: determining whether $c$ is a *unique winner* (that is, $c$ is the only winner), or determining whether $c$ is a *co-winner* (that is, $c$ is in the set of winners).

We consider the setting where the number of alternatives is unbounded and the votes are unweighted. We completely characterize the complexity of possible/necessary winner problems for the following common voting rules: a class of positional scoring rules (including Borda), Copeland, maximin, Bucklin, ranked pairs, voting trees, and plurality with runoff.

## 1. Introduction

In multiagent systems, often, the agents must make a joint decision in spite of the fact that they have different preferences over the alternatives. For example, the agents may have to decide on a joint plan or an allocation of tasks/resources. A general solution to this problem is to have the agents *vote* over the alternatives. That is, each agent $i$ gives a ranking (linear order) $\succ_i$ of all the alternatives; then a *voting rule* takes all of the submitted rankings as input, and based on this produces a chosen alternative (the *winner*), or a set of chosen alternatives. The design of good voting rules has been studied for centuries by the *social choice* community. More recently, computer scientists have become interested in social choice—motivated in part by applications in multiagent systems, but also by other applications. Hence, a community interested in *computational social choice* has emerged.

In "traditional" social choice, agents are usually required to give a linear order over all the alternatives. However, especially in multiagent systems applications, this is not always practical. For one, sometimes, the set of alternatives is too large. For example, there are generally too many possible joint plans or allocations of tasks/resources for an agent to give a linear order over them. In such settings, agents must use a different *voting language* to represent their preferences; for example, they can use CP-nets (Boutilier, Brafman, Domshlak, Hoos, & Poole, 2004; Lang, 2007; Xia, Lang, & Ying, 2007a, 2007b; Lang & Xia, 2009). However, when an agent uses a CP-net (or a similar language) to represent its preferences, this generally only gives us a partial order over the alternatives. Another issue is that it is not always possible for an agent to compare two alternatives (Pini, Rossi, Venable, & Walsh, 2007). Such incomparabilities also result in a partial order.

In this paper, we study the setting where for each agent, we have a partial order corresponding to that agent's preferences. We study the following two questions. (1) Is it the case that, for *some* extension of the partial orders to linear orders, alternative $c$ wins? (2) Is it the case that, for *any* extension of the partial orders to linear orders, alternative $c$ wins? These problems are known as the *possible winner* and *necessary winner* problems, respectively, introduced by Konczak and Lang (2005). Depending on the interpretation of "$c$ wins", the possible/necessary winner problems are further divided into two sub-problems: one is called the possible/neccessary unique winner problem (here "unique" is often omitted when causing no confusion), in which "$c$ wins" means that $c$ is the only winner of the election; the other is called the possible/necessary co-winner problem, in which "$c$ wins" means that $c$ is one of the winners. It should be noted that the answer depends on the voting rule used. Previous research has also investigated the setting where there is uncertainty about the *voting rule*; here, a necessary (possible) winner is an alternative that wins for any (some) realization of the rule (Lang, Pini, Rossi, Venable, & Walsh, 2007). In this paper, we will not study this setting; that is, the rule is always fixed.

While these problems are motivated by the above observations on the impracticality of submitting linear orders, they also relate to *preference elicitation* and *manipulation*. In preference elicitation, the idea is that, instead of having each agent report its preferences all at once, we ask them simple queries about their preferences (*e.g.* "Do you prefer $a$ to $b$?"), until we have enough information to determine the winner. Preference elicitation has found many applications in multiagent systems, especially in combinatorial auctions (for overviews, see Parkes, 2006; Sandholm & Boutilier, 2006) and in voting settings as well (Conitzer & Sandholm, 2002, 2005b; Conitzer, 2009). The problem of deciding whether we can terminate preference elicitation and declare a winner is exactly the necessary winner problem. Manipulation is said to occur when an agent casts a vote that does not correspond to its true preferences, in order to obtain a result that it prefers. By the Gibbard–Satterthwaite Theorem (Gibbard, 1973; Satterthwaite, 1975), for any reasonable voting rule, there are situations where an agent can successfully manipulate the rule. To prevent manipulation, one approach that has been taken in the computational social choice community is to study whether manipulation is (or can be made) computationally hard (Bartholdi, Tovey, & Trick, 1989a; Bartholdi & Orlin, 1991; Hemaspaandra & Hemaspaandra, 2007; Elkind & Lipmaa, 2005; Conitzer, Sandholm, & Lang, 2007; Faliszewski, Hemaspaandra, & Schnoor, 2008; Zuckerman, Procaccia, & Rosenschein, 2009; Xia, Zuckerman, Procaccia, Conitzer, & Rosenschein, 2009; Faliszewski, Hemaspaandra, & Schnoor, 2010). The fundamental questions that have been studied here are "Given the other votes, can a coalition of agents cast their votes so that alternative $c$ wins?" (so-called *constructive manipulation*) and "Given the other votes, can this coalition of agents cast their votes so that alternative $c$ does not win?" (so-called *destructive manipulation*). These problems correspond to the possible winner problem and (the complement of) the necessary winner problem, respectively. To be precise, they only correspond to restricted versions of the possible winner problem and (the complement of) the necessary winner problem in which some of the partial orders are linear orders (the nonmanipulators' votes) and the other partial orders are empty (the manipulators' votes). However, if there is uncertainty about parts of the nonmanipulators' votes, or if parts of the manipulators' votes are already fixed (for example due to preference elicitation), then they can correspond to the general versions of the possible winner problem and (the complement of) the necessary winner problem.

Another related problem is the EVALUATION problem (Conitzer et al., 2007). We are given a probability distribution over each voter's vote, and we are asked for the probability that a given alternative wins. It has been shown that for any anonymous voting rule, when the number of alternatives is no more than a constant, there is a polynomial-time algorithm that solves the EVALUATION problem; when the number of alternatives is not bounded above by a constant, the problem becomes #P hard for plurality, Borda, and Copeland rules (Hazon, Aumann, Kraus, & Wooldridge, 2008). The complexity of influencing the distribution over the voters' votes on multiple binary issues to make a given alternative (a valuation of all these issues) win has also been studied (Erdélyi, Fernau, Goldsmith, Mattei, Raible, & Rothe, 2009). The possible/necessary winner problems are related to the EVALUATION problem in the following way. If every voter assigns positive probability to every one of the linear orders that extend her partial order, then, for any alternative $c$, $c$ is a possible winner if and only if the probability that $c$ wins the election is positive; $c$ is a necessary winner if and only if the probability that $c$ wins the election is 1. We must note that this reduction from the possible/necessary winner problem to the EVALUATION problem is in general not polynomial, because for any partial order, it is possible that there are exponentially many linear orders that extend it. For example, if the partial order is empty, then any linear order is an extension of it. However, in this paper, we prove results that show that the possible/necessary winner problem is hard even when the number of undetermined pairs in each partial order is a constant, so that there are in fact only polynomially many linear orders that extend it. Hence, our hardness results also imply (only NP-)hardness results for the EVALUATION problem.

Because of the variety of different interpretations of the possible and necessary winner problems, it is not surprising that there have already been significant studies of these problems. Two main settings have been studied (see Walsh, 2007 for a good survey). In the first setting, the number of alternatives is bounded, and the votes are weighted. Here, for the Borda, veto, Copeland, maximin, STV, and plurality with runoff rules, the possible winner problem is NP-complete; for the STV and plurality with runoff rules, the necessary winner problem is coNP-complete (Conitzer et al., 2007; Pini et al., 2007; Walsh, 2007). However, in many elections, votes are unweighted (that is, each agent's vote counts the same). If the votes are unweighted, and the number of alternatives is bounded, then the possible and necessary winner problems can always be solved in polynomial time, assuming the voting rule can be executed in polynomial time (Conitzer et al., 2007; Walsh, 2007). Hence, the other setting that has been studied is that where the votes are unweighted and the number of alternatives is not bounded; this is the setting that we will study in this paper. In this setting, the possible and necessary winner problems are known to be hard for STV (Bartholdi & Orlin, 1991; Pini et al., 2007; Walsh, 2007). Computing whether an alternative is a possible or necessary Condorcet winner can be done in polynomial time (Konczak & Lang, 2005). However, at the time of the conference version of this work (Xia & Conitzer, 2008), for most of the other common rules, there were no prior results (except for the fact that the problems are easy for many of these rules when each partial order is either a linear order or empty, that is, the standard manipulation problem).[1]

## 1.1 Our Contributions

In this paper, we characterize the complexity of the possible and necessary winner problems for some of the most important other rules—specifically, a class of positional scoring rules, Copeland, maximin, Bucklin, ranked pairs, voting trees, and plurality with runoff. We show that the possible winner problems are NP-complete for all these rules except the possible unique winner problem with respect to plurality with runoff. We also show that the necessary winner problems are coNP-complete for the Copeland, ranked pairs, and voting trees; and the necessary co-winner problem is coNP-complete for plurality with runoff. For the remaining cases, we present polynomial-time algorithms. Our results are summarized in Table 1.

|  | **Possible Winner** | **Necessary Winner** |
|---|---|---|
| **STV** | NP-complete (Bartholdi & Orlin, 1991) | coNP-complete (Bartholdi & Orlin, 1991) |
| **Plurality** | P [2] | P [2] |
| **Veto** | P [3] | P [3] |
| **Pos. scoring** (incl. **Borda**, $k$-**approval**) | NP-complete [4] | P |
| **Copeland** | NP-complete [4] | coNP-complete [4] |
| **Maximin** | NP-complete [4] | P |
| **Bucklin** | NP-complete [4] | P |
| **Ranked pairs** | NP-complete [4] | coNP-complete [4] |
| **Voting trees** (incl. **balanced trees**) | NP-complete [4] | coNP-complete [4] |
| **Plu. w/ runoff** | NP-complete (unique winner) P (co-winner) | P (unique winner) coNP-complete (co-winner) [4] |

Table 1: Summary of complexity of possible/necessary winner problems with respect to common voting rules. Unless otherwise mentioned, the results do not depend on whether we consider the unique-winner or the co-winner version of the problem.

---

1. An earlier paper (Konczak & Lang, 2005) studied these problems for positional scoring rules, and claimed that the problems are polynomial-time solvable for positional scoring rules; however, there was a subtle mistake in their proofs. We will show that the possible winner problem is in fact NP-complete for some positional scoring rules. We will also give a correct proof that the necessary winner problem is indeed polynomial-time solvable for all positional scoring rules.
2. Easy to prove; also proved in the work of Betzler and Dorn (2010), and follows from the bribery algorithm by Faliszewski (2008).
3. Easy to prove, also proved in the work of Betzler and Dorn (2010).
4. Hardness results hold even when the number of unknown pairs in each partial order is no more than a constant.

This paper is a significant extension of the conference version of this work (Xia & Conitzer, 2008): this extended version includes all the proofs, and the results on voting trees, plurality with runoff, and $k$-approval are new. The conference version also did not mention plurality and veto; these results are easy and follow from known results, as explained in the footnotes under the table.

## 1.2 Subsequent Work since the Conference Version

Since the conference version of this work, the complexity of the possible winner problem with respect to any positional scoring rule has been fully characterized (Betzler & Dorn, 2010; Baumeister & Rothe, 2010). By the theorems of Betzler and Dorn (2010), the possible winner problem is NP-complete with respect to Borda and $k$-approval. Still, these hardness results do not directly imply the hardness results obtained for positional scoring rules in this paper—we prove that the hardness results for Borda and $k$-approval hold even when the number of undetermined pairs in each vote is no more than 4.

Also, a special case of the possible and necessary winner problems where new alternatives join the election after the voters' preferences over the initial alternatives have been fully revealed has been proposed and studied in the work of Chevaleyre, Lang, Maudet, and Monnot (2010). It has been shown that the possible-winner-with-new-alternatives problem is NP-complete for maximin, Copeland (Xia, Lang, & Monnot, 2011), and $k$-approval when $k \geq 3$ and there are at least 3 new alternatives (Chevaleyre, Lang, Maudet, Monnot, & Xia, 2010); the problem is in P for Bucklin (when $k \geq 3$) (Xia, Lang, & Monnot, 2011), Borda, and $k$-approval (when $k \leq 2$ or there are no more than two new alternatives) (Chevaleyre, Lang, Maudet, Monnot, & Xia, 2010).

Meanwhile, a number of new results on the complexity of the unweighted coalitional manipulation problem have also been obtained. Specifically, the unweighted coalitional manipulation problem has been shown to be NP-hard for Copeland$_\alpha$ for any $0 \leq \alpha \leq 1$ (except for $\alpha = \frac{1}{2}$; these results even hold with two manipulators) (Faliszewski et al., 2008, 2010),[5] maximin (two manipulators) and ranked pairs (one manipulator) (Xia et al., 2009), and a specific positional scoring rule (two manipulators) (Xia, Conitzer, & Procaccia, 2010). As we mentioned before, the unweighted coalitional manipulation problem is a special case of the possible winner problem studied in this paper (where some partial orders are linear orders and the others are empty); as a result, NP-hardness results for the unweighted coalitional manipulation problem also imply NP-hardness of the possible winner problem for these rules. We note that the NP-hardness results proved in this paper (except the possible unique winner problem for plurality with runoff) hold even when for each partial order, the number of pairs of alternatives for which the order is unknown is a constant. Therefore, the subsequent research on the unweighted coalitional manipulation does not completely imply the NP-hardness results that we prove in this paper for the possible winner problem for Copeland, maximin, ranked pairs, and positional scoring rules.

Elkind et al. (2009) showed that the possible winner problem also reduces to the *swap bribery* problem, in which an interested party can pay voters to swap adjacent alternatives in their rankings, but the price to swap two alternatives depends on both the identity of the alternatives and the identity of the voter. That is, (with respect to a fixed voting rule) the computational complexity of the swap bribery problem is at least as high as that of the possible winner problem, in terms of polynomial-time reductions.

---

5. Faliszewksi et al. (2008) also study the case of weighted coalitional manipulation with three alternatives for Copeland, and show that how hard this problem is depends both on $\alpha$ and whether we consider the unique-winner or the co-winner variant of the problem. We do not study weighted votes in this paper.

The complexity of the possible winner problem has also been studied from a fixed-parameter tractability perspective, for parameters such as the number of alternatives, the number of voters, and the number of unknown pairs in each vote (Betzler, Hemmann, & Niedermeier, 2009). Finally, the counting version of the possible winner problem has also been studied (Bachrach, Betzler, & Faliszewski, 2010).

## 2. Preliminaries

Let $\mathcal{C} = \{c_1, \ldots, c_m\}$ be the set of *alternatives* (or *candidates*). A linear order on $\mathcal{C}$ is a transitive, antisymmetric, and total relation on $\mathcal{C}$. The set of all linear orders on $\mathcal{C}$ is denoted by $L(\mathcal{C})$. An $n$-voter profile $P$ on $\mathcal{C}$ consists of $n$ linear orders on $\mathcal{C}$. That is, $P = (V_1, \ldots, V_n)$, where for every $i \leq n$, $V_i \in L(\mathcal{C})$. The set of all profiles on $\mathcal{C}$ is denoted by $P(\mathcal{C})$. In the remainder of the paper, $m$ denotes the number of alternatives and $n$ denotes the number of voters.

A *voting rule* $r$ is a function from the set of all profiles on $\mathcal{C}$ to the set of (nonempty) subsets of $\mathcal{C}$, that is, $r : P(\mathcal{C}) \to 2^{\mathcal{C}} \setminus \emptyset$. The following are some common voting rules.

1. *(Positional) scoring rules*: A positional scoring rule is defined by a *scoring vector* $\vec{s}_m = (\vec{s}_m(1), \ldots, \vec{s}_m(m))$ of $m$ non-negative integers, where $\vec{s}_m(1) \geq \cdots \geq \vec{s}_m(m)$. For any vote $V \in L(\mathcal{C})$ and any $c \in \mathcal{C}$, let $s(V, c) = \vec{s}_m(j)$, where $j$ is the rank of $c$ in $V$. For any profile $P = (V_1, \ldots, V_n)$, let $s(P, c) = \sum_{i=1}^{n} s(V_i, c)$. The rule will select $c \in \mathcal{C}$ so that $s(P, c)$ is maximized. Some examples of positional scoring rules are *Borda*, for which the scoring vector is $(m-1, m-2, \ldots, 0)$, *plurality*, for which the scoring vector is $(1, 0, \ldots, 0)$, *veto*, for which the scoring vector is $(1, \ldots, 1, 0)$, and $k$-*approval* $(1 \leq k \leq m-1)$, for which the scoring vector is $(\underbrace{1, \ldots, 1}_{k}, 0, \ldots, 0)$. In this paper, we assume that the scoring vector can be computed in polynomial time.

2. *Copeland*: For any two alternatives $c_i$ and $c_j$, we can simulate a *pairwise election* between them, by seeing how many votes rank $c_i$ ahead of $c_j$, and how many rank $c_j$ ahead of $c_i$. $c_i$ wins if and only if the majority of votes rank $c_i$ ahead of $c_j$. Then, an alternative receives one point for each win in a pairwise election. (Typically, an alternative also receives half a point for each pairwise tie, but this will not matter for our results.) A winner is an alternative who has the highest score.

3. *Maximin (a.k.a. Simpson)*: Let $N_P(c_i, c_j)$ denote the number of votes that rank $c_i$ ahead of $c_j$ in the profile $P$. A winner is an alternative $c$ that maximizes $min\{N_P(c, c') : c' \in \mathcal{C}, c' \neq c\}$.

4. *Bucklin*: An alternative $c$'s Bucklin score is the smallest number $k$ such that more than half of the votes rank $c$ among the top $k$ alternatives. A winner is an alternative who has the smallest Bucklin score. (Sometimes, ties are broken by the number of votes that rank an alternative among the top $k$ position, but for simplicity we will not consider this tiebreaking rule here.)

5. *Ranked pairs*: This rule first creates an entire ranking of all the alternatives. $N_P(c_i, c_j)$ is defined as for the maximin rule. In each step, we will consider a pair of alternatives $c_i, c_j$ that we have not previously considered; specifically, we choose the remaining pair with the highest $N_P(c_i, c_j)$. We then fix the order $c_i > c_j$, unless this contradicts previous orders that we fixed (that is, it violates transitivity). We continue until we have considered all pairs of alternatives (hence we have a full ranking). The alternative at the top of the ranking wins.

6. *Voting trees*: A voting tree is a binary tree with $m$ leaves, where each leaf is associated with an alternative. In each round, there is a pairwise election between an alternative $c_i$ and its sibling $c_j$: if the majority of voters prefer $c_i$ to $c_j$, then $c_j$ is eliminated, and $c_i$ is associated with the parent of these two nodes; similarly, if the majority of voters prefer $c_j$ to $c_i$, then $c_i$ is eliminated, and $c_j$ is associated with the parent of these two nodes. The alternative that is associated with the root of the tree (wins all its rounds) wins. Balanced voting trees are also known as *cup*, *knockout tournaments* or *single-elimination tournaments*.

7. *Plurality with runoff*: The rule has two steps. In the first step, all alternatives except the two that are ranked in the top position for most times are eliminated, and the votes transfer to the second round, in which the plurality rule (a.k.a. *majority* rule in case of two alternatives) is used to select the winner.

8. *Single transferable vote (STV)*: The election has $m$ rounds. In each round, the alternative that gets the minimal plurality score drops out, and is removed from all of the votes (so that votes for this alternative transfer to another alternative in the next round). The last-remaining alternative is the winner.

Given a profile $P$, the *pairwise score difference* $D_P(c, c')$ of alternatives $c$ and $c'$ is defined as follows.

$$D_P(c, c') = N_P(c, c') - N_P(c', c)$$

The subscript $P$ is omitted when there is no risk of confusion. For a linear order $V$ over $\mathcal{C}$, we let $D_V$ denote the pairwise score difference function of the profile that consists of a single vote $V$. That is, $D_V = D_{\{V\}}$. It follows from the definition that $D(c, c') = -D(c', c)$. We note that although maximin, ranked pairs, and voting trees are based on pairwise scores, they can also be computed by pairwise score differences in the same way, because for any profile $P$ of $n$ votes, and any pair of alternatives $(c, c')$, we have $D_P(c, c') = 2N_P(c, c') - n$.

We adopt the *parallel-universes tiebreaking* (Conitzer, Rognlie, & Xia, 2009) to define the winning alternatives for the rules that have multiple rounds (i.e., ranked pairs, voting trees, plurality with runoff, and STV). That is, an alternative $c$ is a winner if and only if there exists a way to break ties in all of the steps such that $c$ is the winner. For example, an alternative $c$ is a winner for a voting tree, if there exists a way to break ties in the pairwise elections in the voting process, such that $c$ wins. A partial order on $\mathcal{C}$ is a reflexive, transitive, and antisymmetric relation on $\mathcal{C}$. We say a linear order $V$ extends a partial order $O$ if $O \subseteq V$.

**Definition 1** *A linear order $V$ on $\mathcal{C}$ extends a partial order $O$ on $\mathcal{C}$ if for every pair of alternatives $c, c' \in \mathcal{C}$, $c \succ_O c' \Rightarrow c \succ_V c'$.*

Throughout the paper we use the following notation. Let $V$ denote a linear order over $\mathcal{C}$; let $O$ denote a partial order over $\mathcal{C}$; let $P$ denote a profile of linear orders; let $P_{poset}$ denote a profile of partial orders.

## 3. Possible/Necessary Winners

We are now ready to define possible (necessary) winners, which were first introduced by Konczak and Lang (2005).

**Definition 2** *Given a profile of partial orders $P_{poset} = (O_1, \ldots, O_n)$ on $\mathcal{C}$, we say that an alternative $c \in \mathcal{C}$ is: (1) a* possible winner *if there exists $P = (V_1, \ldots, V_n)$ such that each $V_i$ extends $O_i$, and $r(P) = \{c\}$; (2) a* necessary winner *if for every $P = (V_1, \ldots, V_n)$ such that each $V_i$ extends $O_i$, $r(P) = \{c\}$; (3) a* possible co-winner *if there exists $P = (V_1, \ldots, V_n)$ such that each $V_i$ extends $O_i$, and $c \in r(P)$; (4) a* necessary co-winner *if for any $P = (V_1, \ldots, V_n)$ such that each $V_i$ extends $O_i$, $c \in r(P)$.*

**Example 1** *Let there be three alternatives $\{c_1, c_2, c_3\}$. Three partial orders are illustrated in Figure 1. Let $P_{poset} = (O_1, O_2, O_3)$. $c_1$ is a possible (co-)winner of $P_{poset}$ with respect to plurality, because we can complete $O_1$ by adding $c_2 \succ c_3$, complete $O_2$ by adding $c_1 \succ c_2$, and complete $O_3$ by adding $c_1 \succ c_2$ and $c_1 \succ c_3$; then, $c_1$ is the only winner. However, $c_1$ is not a necessary (co-)winner, because we can complete $O_1$ by adding $c_2 \succ c_3$, complete $O_2$ by adding $c_2 \succ c_1$, and complete $O_3$ by adding $c_2 \succ c_1$ and $c_1 \succ c_3$; then, $c_2$ is the only winner.*

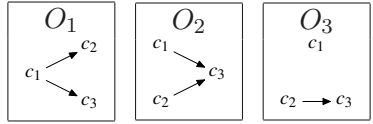

Figure 1: Partial orders.

*However, if we let $P'_{poset} = (O_1, O_1, O_2)$, then $c_1$ is the (only) necessary winner, because $c_1$ will be ranked first in at least two votes.*

Now, we define the computational problems studied in this paper:

**Definition 3** *Define the problem* Possible Winner (PW) *with respect to voting rule $r$ to be: given a profile $P_{poset}$ of partial orders and an alternative $c$, we are asked whether or not $c$ is a possible winner for $P_{poset}$ with respect to $r$.*

*Necessary Winner (NW), Possible co-Winner (PcW)*, and *Necessary co-Winner (NcW)* are defined similarly.

A natural first question is how these problems are related to each other. It turns out that (holding the voting rule fixed) there exists a polynomial-time Turing reduction from NW to PcW. That is, if PcW is in P, then NW is also in P.

**Proposition 1** *For any voting rule $r$, if computing PcW with respect to $r$ is in P, then computing NW with respect to $r$ is also in P.*

*Proof.* Because $r$ never outputs $\emptyset$, an alternative $c$ is a necessary unique winner with respect to $r$ if and only if for every alternative $d$ ($d \neq c$), $d$ is not a possible co-winner. Therefore, if we have a polynomial-time algorithm that solves the PcW problem with respect to $r$, then to solve the NW problem, we simply run the algorithm for every alternative $d$ ($d \neq c$). If any $d \neq c$ is a possible co-winner, then we output that $c$ is not a necessary unique winner; otherwise, we output that $c$ is a necessary unique winner. $\square$

There is no similar relationship between the PW and NcW problems. It is true that if some alternative $d$ ($d \neq c$) is a possible unique winner, then $c$ is not a necessary co-winner. However, it is possible that even if no alternative $d$ ($d \neq c$) is a possible unique winner, $c$ is still not a necessary co-winner. For example, let $P_{poset} = (c_2 \succ c_3 \succ c_1, c_3 \succ c_2 \succ c_1)$. Because $P_{poset}$ is already composed of linear orders, it has only one extension (itself). It follows that there is no possible unique winner for $P_{poset}$ with respect to plurality, but clearly $c_1$ is not a necessary co-winner. More generally, we have the following proposition, which says that for any pair of different problems $X, Y \in \{\text{PW, PcW, NW, NcW}\}$, the answer to $Y$ cannot be computed from only the answers to X for all alternatives, unless X =NW and Y =PcW. (This holds even when the rule is plurality).

**Proposition 2** *Suppose $m \geq 3$. Let $X, Y \in \{PW, PcW, NW, NcW\}$ be such that (1) $X \neq Y$ and (2) $X \neq PcW$ or $Y \neq NW$. There exist two profiles $P_{poset}$ and $\bar{P}_{poset}$ of partial orders (with $|P_{poset}| = |\bar{P}_{poset}|$), such that (1) for every alternative $c$, the answers to X with respect to plurality are the same for both $P_{poset}$ and $\bar{P}_{poset}$, and (2) there exists an alternative $d$ for which the answers to Y with respect to plurality for $P_{poset}$ and $\bar{P}_{poset}$ are different.*

*Proof.* The proof is by construction. For any partial order $O$ on $\mathcal{C}$, we let $Top(O)$ denote the set of alternatives $c'$ for which there exists at least one extension of $O$ where $c'$ is in the top position. For any set $\mathcal{C}' \subseteq \mathcal{C}$, we define $O_{\mathcal{C}'}$ to be an arbitrary partial order such that $Top(O_{\mathcal{C}'}) = \mathcal{C}'$. For simplicity, we write $O_{c'}$ for $O_{\{c'\}}$. For example, when $m = 3$, we can let $O_{c_1} = O_{\{c_1\}} = c_1 \succ c_2 \succ c_3$, that is, $O_{c_1}$ is a linear order. As another example, $O_{\{c_1, c_2\}}$ is the partial order that is obtained from $[c_1 \succ c_2 \succ c_3]$ by removing $c_1 \succ c_2$.

Let $d = c_1$. We next specify the profiles $P_{poset}$ and $\bar{P}_{poset}$ for the following (exhaustive) list of cases $(X, Y)$.

- (PW, NW) and (PW, NcW). (1) Let $P_{poset}$ be composed of 5 copies of $O_{c_1}$. $c_1$ is a necessary unique/co-winner. (2) Let $\bar{P}_{poset} = (O_{c_1}, O_{c_2}, O_{c_2}, O_{\{c_1, c_3\}}, O_{\{c_1, c_3\}})$. $c_1$ is the unique winner for one extension, $\{c_1, c_2\}$ are the winners for two extensions, and $\{c_2, c_3\}$ are the winners for one extension. Therefore, $c_1$ is not a necessary unique/co-winner. We note that $c_1$ is the only possible unique winner for both profiles.

- (PW, PcW). (1) Let $P_{poset} = (O_{c_1}, O_{c_2})$. $\{c_1, c_2\}$ are the winners for the only extension, which means that $c_1$ is a possible co-winner. (2) Let $\bar{P}_{poset} = (O_{c_2}, O_{c_3})$. $\{c_2, c_3\}$ are the winners for the only extension, which means that $c_1$ is not a possible co-winner. We note that there is no possible unique winner for either profile.

- (NcW, NW). (1) Let $P_{poset} = (O_{c_1}, O_{c_1})$. $c_1$ is the only winner in the only extension, which means that $c_1$ is the necessary unique winner. (2) Let $\bar{P}_{poset} = (O_{c_1}, O_{\{c_1, c_2\}})$. $c_1$ is the only winner for one extension, and $\{c_1, c_2\}$ are the winners for the other extension, which means that $c_1$ is not the necessary unique winner. We note that $c_1$ is the only necessary co-winner for both profiles.

- (PcW, PW) and (PcW, NcW). (1) Let $P_{poset} = (O_{\{c_1,c_2\}}, O_{\{c_1,c_2\}})$. $c_1$ is the unique winner for one extension, $c_2$ is the unique winner for one extension, and $\{c_1, c_2\}$ are the winners for two extensions. Therefore, $c_1$ is a possible unique winner (meanwhile, $c_1$ is not a necessary co-winner). (2) Let $\bar{P}_{poset} = (O_{c_1}, O_{c_2})$. $\{c_1, c_2\}$ are the winners for the only extension, which means that $c_1$ is not a possible unique winner (meanwhile, $c_1$ is a necessary co-winner). We note that $c_1$ and $c_2$ are the only possible co-winners for both profiles.

- (NW, PW), (NW, PcW), (NcW, PW), (NcW, PcW). (1) Let $P_{poset} = (O_{\{c_1,c_2\}}, O_{\{c_1,c_2\}})$. $c_1$ is the unique winner for one extension, $c_2$ is the unique winner for one extension, and $\{c_1, c_2\}$ are the winners for two extensions. Therefore, $c_1$ is a possible unique/co-winner. (2) Let $\bar{P}_{poset} = (O_{\{c_2,c_3\}}, O_{\{c_2,c_3\}})$. $c_2$ is the unique winner for one extension, $c_3$ is the unique winner for one extension, and $\{c_2, c_3\}$ are the winners for two extensions. Therefore, $c_1$ is not a possible unique/co-winner. We note that there is no necessary unique/co-winner for either profile.

- (NW, NcW). (1) Let $P_{poset} = (O_{c_1}, O_{c_2})$. $\{c_1, c_2\}$ are the winners for the only extension, which means that $c_1$ is a necessary co-winner. (2) Let $\bar{P}_{poset} = (O_{c_2}, O_{\{c_1,c_2\}})$. $\{c_1, c_2\}$ are the winners for one extension, and $c_2$ is the unique winner for the other extension, which means that $c_1$ is not a necessary co-winner. We note that there is no necessary unique winner for either profile.

□

## 4. Hardness Results

In this section, we prove that PW (PcW) is NP-complete with respect to a class of positional scoring rules, Copeland, maximin, Bucklin, ranked pairs, and voting trees; NW (NcW) is coNP-complete with respect to Copeland, ranked pairs, and voting trees; and PW is NP-complete and NcW is coNP-complete with respect to plurality with runoff. For positional scoring rules, we will not show that PW is hard for all positional scoring rules—in fact, for plurality and veto, PW is easy; rather, we will give a sufficient condition on a positional scoring rule under which PW is hard. Most notably, Borda satisfies this condition. $k$-approval does not satisfy this condition, and we will provide a distinct proof for PW (PcW) with respect to $k$-approval ($k \geq 2$).[6] Similarly for voting trees, we provide a necessary condition under which the hardness results hold, and most notably, balanced voting trees satisfy this condition. All of these results (except the one for PW with respect to plurality with runoff) hold even when the partial orders are "almost" linear orders. That is, the number of undetermined pairs in each partial order is bounded above by a constant.

---

6. After the conference version of this paper (Xia & Conitzer, 2008), Betzler and Dorn proved a dichotomy theorem for possible winner problems with respect to positional scoring rules (Betzler & Dorn, 2010). According to their theorem, PW with respect to $k$-approval ($k \geq 2$) is NP-complete. In this paper, we prove that the problem is NP-complete, even when the number of undetermined pairs in each vote is no more than 4.

All the hardness results are proved by reductions from the EXACT 3-COVER (X3C) problem, except for the result for $k$-approval, which is proved by a reduction from 3-SAT. X3C and 3-SAT are known to be NP-complete (Garey & Johnson, 1979). The two problems are defined as follows.

**Definition 4 (X3C)** *We are given a set $\mathcal{V} = \{v_1, \ldots, v_q\}$ and a collection $\mathcal{S} = \{S_1, \ldots, S_t\}$, where for each $i \leq t$, $S_i = \{v_{l(i,1)}, v_{l(i,2)}, v_{l(i,3)}\} \subseteq \mathcal{V}$, with $1 \leq l(i,1), l(i,2), l(i,3) \leq q$. We are asked whether we can cover all of the elements in $\mathcal{V}$ with non-overlapping sets in $\mathcal{S}$.*

**Definition 5 (3-SAT)** *We are given a formula in conjunctive normal form: $F = C_1 \wedge \ldots \wedge C_t$ over binary variables $\mathbf{x}_1, \ldots, \mathbf{x}_q$, where for any $j \leq t$, $C_j$ is called a* clause. *For any $j \leq t$, $C_j = l_j^1 \vee l_j^2 \vee l_j^3$, where for any $\alpha \in \{1, 2, 3\}$, $l_j^\alpha$ is called a* literal, *and there exists $i \leq q$ such that either $l_j^\alpha = \mathbf{x}_i$ or $l_j^\alpha = \neg \mathbf{x}_i$. We are asked whether there exists a valuation of the variables under which $F$ is true.*

In each proof, the election instance that we construct from an arbitrary X3C (or 3-SAT) instance consists of two parts. The first part is a set of partial orders that encode the X3C (or 3-SAT) instance.[7] For example, in some of our PW reductions from X3C, the first part is structured as follows: in order for $c$ to win, there is an alternative $c'$ that needs to be placed in a "high" position in the extensions of the partial orders at least some number of times. However, for each of the partial orders, there is a set of three alternatives such that if we put $c'$ in a high position in an extension of that partial order, then these three alternatives must be ranked in even higher positions (that is, $c'$ "pushes up" these three alternatives in the extension). These sets of three alternatives that must sometimes be pushed up correspond to the sets of three elements in the X3C instance. The PW instance is set up in such a way that if the same X3C-element alternative is pushed up by $c'$ in two different votes in the first part, then $c$ cannot win. Thus, the sets of alternatives that we push up must be disjoint, and the instance is set up in such a way that we need to put $c'$ in a high position often enough that the pushed-up 3-sets actually must constitute an exact cover. The second part is a set of linear orders (that is, in the second part, everything is determined) whose purpose is, informally stated, to adjust the scores of the alternatives so that we get the properties just described.

First we introduce some notation to represent the set of all pairwise comparisons in a linear order.

**Definition 6** *For any set $\{a_1, \ldots, a_l\}$, let $O(a_1, \ldots, a_l) = \{(a_i, a_j) : i < j\}$.*

That is, $O(a_1, \ldots, a_l)$ is the set of all ordered pairs consistent with the linear order $a_1 \succ \ldots \succ a_l$. For example, $O(a, b, c) = \{(a, b), (b, c), (a, c)\}$. The following notation will be frequently used in the proofs.

**Definition 7** *For any set $A$ and any partition $A_1, \ldots, A_k$ of $A$, let $O(A_1, \ldots, A_k)$ denote an arbitrary linear order on $A$ that is consistent with $A_1 \succ A_2 \succ \ldots \succ A_k$.*

---

7. Typically, we define the partial orders by first defining some linear orders and then removing some of the pairwise ordering constraints.

The proofs that make use of this notation only use the fact that $O(A_1, \ldots, A_k)$ is consistent with $A_1 \succ \ldots \succ A_k$, so that the order within each $A_i$ ($i \leq k$) does not matter. For example, let $A = \{a, b, c, d\}$, $A_1 = \{a\}$, $A_2 = \{b, c\}$, $A_3 = \{d\}$. There are two linear orders that are consistent with $A_1 \succ A_2 \succ A_3$. They are $a \succ b \succ c \succ d$ and $a \succ c \succ b \succ d$. $O(A_1, A_2, A_3)$ can denote either of them, e.g., $O(A_1, A_2, A_3) = a \succ b \succ c \succ d$. Sometimes we use the notation "Others" to denote the set of all objects that are not mentioned in the context. For example, $O(A_1, A_2, A_3) = O(\text{Others}, A_2, A_3) = O(A_1, \text{Others}, A_3) = O(A_1, A_2, \text{Others})$.

Usually, a positional scoring rule is defined for a fixed number of alternatives (that is, $m$ is fixed). If we hold $m$ fixed, then there exist polynomial-time algorithms for both PW and NW (Walsh, 2007; Conitzer et al., 2007). However, there are positional scoring rules that are defined for any number of alternatives—for example, Borda, plurality, and veto. For such positional scoring rules, the number of alternatives is not bounded, and indeed, we will prove that PW is not always easy with respect to such rules. To study the complexity of social choice problems that involve a growing number of alternatives, it is necessary to associate a scoring vector with every natural number of alternatives. In the remainder of the paper, a positional scoring rule $r$ consists of a sequence of scoring vectors $\{\vec{s}_1, \vec{s}_2, \ldots\}$ such that for each $i \in \mathbb{N}$, $\vec{s}_i$ is the scoring vector for $i$ alternatives. The next theorem provides a sufficient condition on a positional scoring rule for PW to be NP-complete. In this paper, all the PW/PcW problems are in NP, and all the NW/NcW problems are in coNP. This follows from the fact that, given an extension of the partial orders to linear orders, we can compute the winner(s) in polynomial-time for the rules studied in this paper. With this in mind, we only prove the hardness direction in the NP-completeness/coNP-completeness proofs. There do exist rules for which computing the winner(s) is NP-hard, for example, Dodgson's rule (Bartholdi, Tovey, & Trick, 1989b; Hemaspaandra, Hemaspaandra, & Rothe, 1997) and Young's rule (Rothe, Spakowski, & Vogel, 2003), but we will not study any rules for which computing the winners is hard here.

**Theorem 1** *Let $r$ be a positional scoring rule with scoring vectors $\{\vec{s}_1, \vec{s}_2, \ldots\}$. Suppose there exists a polynomial function $f(x)$ such that for any $x \in \mathbb{N}$, there exist $l$ and $k$ with $x \leq l \leq f(x)$ and $k \leq l - 4$, and satisfy the following conditions:*

*(1) $\vec{s}_l(k) - \vec{s}_l(k+1) = \vec{s}_l(k+1) - \vec{s}_l(k+2) = \vec{s}_l(k+2) - \vec{s}_l(k+3) > 0$,*

*(2) $\vec{s}_l(k+3) - \vec{s}_l(k+4) > 0$,*

*Then, PW and PcW are both NP-complete with respect to $r$, even when the number of undetermined pairs in each vote is no more than 4.*

*Proof.* Given an X3C instance $\mathcal{V} = \{v_1, \ldots, v_q\}$, $\mathcal{S} = \{S_1, \ldots, S_t\}$, let $q + 3 \leq l \leq f(q+3)$ (where $q$ is the number of elements in the X3C instance) satisfy the two conditions in the assumption, and let $k \leq l - 4$ satisfy $\vec{s}_l(k) - \vec{s}_l(k+1) = \vec{s}_l(k+1) - \vec{s}_l(k+2) = \vec{s}_l(k+2) - \vec{s}_l(k+3) > 0$, and $\vec{s}_l(k+3) - \vec{s}_l(k+4) > 0$. Let $K_1 = \vec{s}_l(k) - \vec{s}_l(k+1)$ and $K_2 = \vec{s}_l(k+3) - \vec{s}_l(k+4)$. We construct the PW instance as follows.
**Alternatives:** $\mathcal{C} = \{c, w, d\} \cup \mathcal{V} \cup A$, where $d$ and $A = \{a_1, \ldots, a_{l-q-3}\}$ are auxiliary alternatives.
**First part ($P_1$) of the profile:** For each $j \leq t$, choose an arbitrary set $B_j \subset \mathcal{C} \setminus (S_i \cup \{w, d\})$

DETERMINING POSSIBLE AND NECESSARY WINNERS GIVEN PARTIAL ORDERSwith $|B_j| = k - 1$. We define a partial order $O_j$ as follows.

$$O_j = O(B_j, w, S_i, d, \text{Others}) \setminus [\{w\} \times (S_j \cup \{d\})]$$

That is, $O_j$ is a partial order that agrees with $B_j \succ w \succ S_j \succ d \succ$ Others, except that the pairwise relations between $(w, S_j)$ and $(w, d)$ are not determined (and these are the only 4 undetermined relations). Let $P_1 = \{O_1, \ldots, O_t\}$.

**Second part ($P_2$) of the profile:** We first give the properties that we need $P_2$ to satisfy; we will show how to construct $P_2$ in polynomial time later in the proof. All votes in $P_2$ are linear orders. Let $P_1' = \{O(B_j, w, S_j, d, \text{Others}) : j \leq t\}$. That is, $P_1'$ ($|P_1'| = t$) is an extension of $P_1$ (in fact, $P_1'$ is the set of linear orders that we started with to obtain $P_1$, before removing some of the pairwise relations). $P_2$ is a set of linear orders such that the following holds for $Q = P_1' \cup P_2$:

(1) For every $i \leq q$, $\vec{s}_l(Q, c) - \vec{s}_l(Q, v_i) = 2K_1$, $\vec{s}_l(Q, w) - \vec{s}_l(Q, c) = \frac{q}{3} \times (3K_1 + K_2) - K_2$.

(2) For every $i \leq q$, the scores of $v_i$ and $w, c$ are higher than those of the other alternatives in any extension of $P_1 \cup P_2$.

(3) $P_2$'s size is polynomial in $t + q$.

Suppose there exists an extension $P_1^*$ of $P_1$ such that $c$ is the winner for $P_1^* \cup P_2$. For each $i \leq q$, $v_i$ is not ranked higher than $w$ more than once in $P_1^*$, because otherwise the total score of $v_i$ will be higher than or equal to the total score of $c$. We recall that the score difference between $w$ and $c$ in $P_1' \cup P_2$ is $\frac{q}{3} \times (3K_1 + K_2) - K_2$. Therefore, if there exists $j \leq t$ such that in the extension of $O_j$, $w$ is ranked above $c$, and is ranked below some alternative in $S_j$, then there must exist an alternative in $\mathcal{V}$ that is ranked above $w$ at least two times in $P_1^*$, which contradicts the assumption that $c$ is the winner. It follows that in order for the total score of $w$ to be lower than the total score of $c$, $w$ is ranked lower than $d$ at least $\frac{q}{3}$ times. Let $I$ denote the set of subscripts of votes in $P_1^*$ for which $w$ is ranked lower than $d$; then, $S_I = \{S_i : i \in I\}$ is a solution to the X3C instance.

Conversely, given a solution to the X3C instance, let $I$ be the set of indices of $S_i$ that are included in the X3C. Then, a solution to the possible winner instance can be obtained by ranking $d$ ahead of $w$ exactly in the votes with subscripts in $I$. Therefore, $c$ is a possible winner if and only if there exists a solution to the X3C problem, which means that PW and PcW are NP-complete with respect to positional scoring rules that satisfy the conditions stated in the theorem.

For possible co-winner, we replace (1) by the following condition.
(1') For every $i \leq q$, $s(Q, c) - s(Q, v_i) = K_1$, $s(Q, w) - s(Q, c) = \frac{q}{3} \times (3K_1 + K_2)$.

Next, we show how to construct the profile $P_2$ so that it satisfies the three conditions. $P_2$ consists of the following three parts.

> **The first part, $P_2'$.** Let $M_V$ denote the cyclic permutation among $\mathcal{V} \cup \{c, w\}$. That is, $M_V = c \to w \to v_1 \to v_2 \to \ldots \to v_q \to c$. For any $j \in \mathbb{N}$, and any $e \in \mathcal{V} \cup \{c, w\}$, we let $M_V^0(e) = e$, and $M_V^j(e) = M_V(M_V^{j-1}(e))$. The first part of $P_2$ is $P_2' = M_V(P_1') \cup M_V^2(P_1') \cup \ldots \cup M_V^{q+1}(P_1')$. It follows that for any $e, e' \in \mathcal{V} \cup \{c, w\}$, $\vec{s}_l(P_1' \cup P_2', e) = \vec{s}_l(P_1' \cup P_2', e')$.

**The second part, $P_2^*$.** Choose an arbitrary set $B \subseteq \mathcal{C} \setminus \{d, w, c\}$ such that $|B| = k-1$, and an arbitrary set $A' \subseteq \mathcal{C} \setminus (B \cup \{d, w\})$ such that $|A'| = 3$. We define the following partial orders.

$$\begin{aligned}
V_1 &= O(B, d, w, c, \text{Others}), & V_1' &= O(B, c, w, d, \text{Others}) \\
V_2 &= O(B, d, c, w, \text{Others}), & V_2' &= O(B, w, c, d, \text{Others}) \\
V_3 &= O(B, d, A', w, \text{Others}), & V_3' &= O(B, w, A', d, \text{Others}) \\
V_4 &= O(B, A', d, w, \text{Others}), & V_4' &= O(B, A', w, d, \text{Others})
\end{aligned}$$

$P_2^*$ is defined as follows.

$$\begin{aligned}
P_2^* = & \{V_1', V_2', M_V(V_1), M_V(V_2), \ldots, M_V^{q+1}(V_1), M_V^{q+1}(V_2)\} \\
& \cup \frac{q}{3} \times \{V_3', M_V(V_3), \ldots, M_V^{q+1}(V_3)\} \cup \{V_4', M_V(V_4), \ldots, M_V^{q+1}(V_4)\}
\end{aligned}$$

Here $\frac{q}{3} \times \{V_3', M_V(V_3), \ldots, M_V^{q+1}(V_3)\}$ represents $\frac{q}{3}$ copies of $\{V_3', M_V(V_3), \ldots, M_V^{q+1}(V_3)\}$. Putting $P_2'$ and $P_2^*$ together, the condition (1) in the description of $P_2$ is satisfied.

**The third part, $\tilde{P}_2$.** $\tilde{P}_2$ is defined in a way such that in $\tilde{P}_2$, the total scores of each pair of alternatives in $\mathcal{V} \cup \{c, w\}$ are the same, and the total score of any alternative in $\mathcal{V} \cup \{c, w\}$ is significantly higher than the total score of any alternative in $A \cup \{d\}$. Let $M_O$ be a cyclic permutation among $A \cup \{d\}$. That is, we let $M_O = d \to a_1 \to a_2 \to \ldots \to a_{l-q-3} \to d$. Let $V_5 = O(\mathcal{V}, c, w, \text{Others})$. We define the third part $\tilde{P}_2$ as follows.

$$\tilde{P}_2 = (|P_1 \cup P_2' \cup P_2^*| + 1) \times \{M_V^i(M_O^j(V_5)) : i \leq q+2, j \leq l-q-2\}$$

We note that $|P_1 \cup P_2' \cup P_2^*| + 1$ is polynomial in $t + q$. Therefore, the size of $\tilde{P}_2$ is polynomial in $t + q$.

$\square$

Theorem 1 provides a sufficient condition on positional scoring rules for PW and PcW to be NP-complete. It can be applied to prove NP-completeness of PW and PcW for Borda, as the following corollary shows.

**Corollary 1** *PW and PcW are NP-complete with respect to Borda, even when the number of undetermined pairs in each vote is no more than 4.*

*Proof.* For any $l \in \mathbb{N}$, the scoring vector $\vec{s}_l$ for Borda is $(l-1, l-2, \ldots, 0)$. If we let $f(x) = x$, $l = x$, and $k = l - 4$, then the conditions in Theorem 1 are all satisfied, and the claim follows. $\square$

Theorem 1 does not apply to $k$-approval. As we noted in Table 1, the possible and necessary winner problems with respect to plurality (1-approval) are in P. We next show that for any fixed $k \in \mathbb{N}$ with $k \geq 2$, PW and PcW with respect to $k$-approval are NP-complete.

**Theorem 2** *For any fixed natural number $k \geq 2$, PW and PcW are NP-complete with respect to $k$-approval, even when the number of undetermined pairs in each vote is no more than 4.*

*Proof.* We first prove the NP-hardness for PW with respect to 2-approval. Then, we show how to extend the proof to any $k \in \mathbb{N}$, where $k \geq 2$.

We prove the NP-hardness by a reduction from 3-SAT. Given an instance of 3-SAT, where there are $q$ variables $\mathbf{x}_1, \ldots, \mathbf{x}_q$ and a formula $F = C_1 \wedge \ldots \wedge C_t$, we construct an instance of PW with respect to 2-approval as follows. Without loss of generality, we assume that $q + t \geq 2$ (generally, for any fixed $k \in \mathbb{N}$, we can assume that $q + t \geq k$), and that in each clause of $F$, no variable appears more than once.

**Alternatives:** $\mathcal{C} = \{c\} \cup C \cup X \cup X_1 \cup X_1^{\neg} \cup \ldots \cup X_q \cup X_q^{\neg} \cup D_1 \cup D_1^{\neg} \cup \ldots \cup D_q \cup D_q^{\neg}$, where $C = \{c_1, \ldots, c_t\}$, $X = \{x_1, \ldots, x_q, \neg x_1, \ldots, \neg x_q\}$, and for each $i \leq q$,

- $X_i = \{x_i^1, \ldots, x_i^t, \hat{x}_i^1, \ldots, \hat{x}_i^t\}$, $X_i^{\neg} = \{\neg x_i^1, \ldots, \neg x_i^t, \neg \hat{x}_i^1, \ldots, \neg \hat{x}_i^t\}$;
- $D_i = \{d_i^1, \ldots, d_i^t\}$, $D_i^{\neg} = \{\neg d_i^1, \ldots, \neg d_i^t\}$.

In words, $C$ represents the set of clauses in $F$; $x_i$ and $\neg x_i$ represent the values that the Boolean variable $\mathbf{x}_i$ can take; $X_i$ (respectively, $X_i^{\neg}$) represents a set of "duplicates" of $x_i$ (respectively, $\neg x_i$); $D_i$ (respectively, $D_i^{\neg}$) represents a set of auxiliary alternatives that are associated with $x_i$ (respectively, $\neg x_i$).

**First part $P_1$ of the profile:** For each $i \leq q$, we let $V_i = O(c, x_i, \neg x_i, \text{Others})$. Then, we obtain $O_i$ by removing $(x_i, \neg x_i)$ from $V_i$. That is, in any extension of $O_i$, $c$ must be in the top position, and one of $x_i$ and $\neg x_i$ must be in the second position (and the other, in the third). We will see later in the proof that the two extensions of $O_i$ correspond to the two valuations of the variable $\mathbf{x}_i$, i.e., $x_i$ being ranked in the second position (while $\neg x_i$ is ranked in the third position) corresponds to $\mathbf{x}_i = \mathit{false}$.

For each $i \leq q$, we define the following linear orders.

$$V_i^1 = O(x_i, d_i^1, x_i^1, \hat{x}_i^1, \text{Others})$$

$$\forall 2 \leq j \leq t, V_i^j = O(\hat{x}_i^{j-1}, d_i^j, x_i^j, \hat{x}_i^j, \text{Others})$$

Then, we obtain $O_i^1$ from $V_i^1$ by removing $\{x_i, d_i^1\} \times \{x_i^1, \hat{x}_i^1\}$; for each $2 \leq j \leq t$, we obtain $O_i^j$ from $V_i^j$ by removing $\{\hat{x}_i^{j-1}, d_i^j\} \times \{x_i^j, \hat{x}_i^j\}$. We define $V_i^{j,\neg}$ and $O_i^{j,\neg}$ similarly by adding $\neg$ to each alternative explicitly written in the definition of $V_i^j$ and $O_i^j$, respectively (that is, the alternatives that are not in "Others"). For example, $V_i^{1,\neg} = O(\neg x_i, \neg d_i^1, \neg x_i^1, \neg \hat{x}_i^1, \text{Others})$.

For each $j \leq t$, let $f_j : X \to X_1 \cup X_1^{\neg} \cup \ldots \cup X_q \cup X_q^{\neg}$ be the mapping such that for any $x \in X$, $f_j(x)$ is obtained from $x$ by adding $j$ to the superscript of $x$. For example, $f_j(x_1) = x_1^j$ and $f_j(\neg x_2) = \neg x_2^j$. For each $j \leq t$, let $W_j = O(c, f_j(l_j^1), f_j(l_j^2), f_j(l_j^3), \text{Others})$. Then, we obtain $Q_j$ from $W_j$ by removing

$$\{f_j(l_j^1), f_j(l_j^2), f_j(l_j^3)\} \times \{f_j(l_j^1), f_j(l_j^2), f_j(l_j^3)\}$$

That is, in any extension of $Q_j$, $c$ must be in the top position, and one of $\{f_j(l_j^1), f_j(l_j^2), f_j(l_j^3)\}$ must be in the second position. We will see that the extensions of $Q_j$ correspond to how $C_j$ (the $j$th clause) is satisfied under a valuation of $\mathbf{x}_1, \ldots, \mathbf{x}_q$.

We let $P_1 = \{O_1, \ldots, O_q\} \cup \{O_i^j, O_i^{j,\neg} : \forall i \leq q, j \leq t\} \cup \{Q_j : \forall j \leq t\}$.

**Second part $P_2$ of the profile:** for any profile $P$ and any alternative $c'$, we let $s^2(P, c')$ denote the score of $c'$ in $P$, under 2-approval. That is, $s^2(P, c')$ is the number of times that $c'$ is ranked in the top two positions in $P$. We let $P_2$ be an arbitrary profile of linear orders that satisfies the following conditions.

- $s^2(P_2, c) = 0$.
- For every $i \leq q$ and every $j \leq t$, $s^2(P_2, x_i) = s^2(P_2, \neg x_i) = s^2(P_2, x_i^j) = s^2(P_2, \neg x_i^j) = s^2(P_2, \hat{x}_i^j) = s^2(P_2, \neg \hat{x}_i^j) = q + t - 2$.
- For any $c'$ not mentioned above, $s^2(P_2, c') \leq 1$.

Because $t + q \geq 2$, $P_2$ is well-defined and $|P_2|$ is bounded above by a polynomial of $t$ and $q$ (we try to fit $q + t - 2$ copies of $\{x_i, \neg x_i, x_i^j, \neg x_i^j, \hat{x}_i^j, \neg \hat{x}_i^j : \forall i \leq q, j \leq t\}$ into the top two positions of $\lceil (q+t-2)(2q+4qt)/2 \rceil = q(q+t-2)(2t+1)$ votes). We note that the number of undetermined pairs in each vote in $P_1 \cup P_2$ is no more than 4.

Suppose there is a feasible solution to the 3-SAT instance. Let $g$ denote a valuation of $\mathbf{x}_1, \ldots, \mathbf{x}_q$ under which $F$ is satisfied. We define an extension of $P_1 \cup P_2$ as follows.

- For every $i \leq q$, if $g(\mathbf{x}_i) = true$, then we define the following extensions of partial orders in $P_1$.
    - Let $\bar{V}_i$ be the extension of $O_i$ in which $\neg x_i$ is ranked in the second position.
    - Let $\bar{V}_i^1$ be an extension of $O_i^1$ in which $x_i$ and $d_i^1$ are ranked in the top two positions; let $\bar{V}_i^{1,\neg}$ be an extension of $O_i^{1,\neg}$ in which $\neg x_i^1$ and $\neg \hat{x}_i^1$ are ranked in the top two positions.
    - For every $2 \leq j \leq t$, let $\bar{V}_i^j$ be an extension of $O_i^j$ in which $\hat{x}_i^{j-1}$ and $d_i^j$ are ranked in the top two positions.
    - For every $2 \leq j \leq t$, let $\bar{V}_i^{j,\neg}$ be an extension of $O_i^{j,\neg}$ in which $\neg x_i^j$ and $\neg \hat{x}_i^j$ are ranked in the top two positions.

- For every $i \leq q$, if $g(\mathbf{x}_i) = false$, then we define the following extensions (which are similar to the extensions in the case where $g(\mathbf{x}_i) = true$).
    - Let $\bar{V}_i$ be the extension of $O_i$ in which $x_i$ is ranked in the second position.
    - Let $\bar{V}_i^{1,\neg}$ be an extension of $O_i^{1,\neg}$ in which $\neg x_i$ and $\neg d_i^1$ are ranked in the top two positions; let $\bar{V}_i^1$ be an extension of $O_i^1$ in which $x_i^1$ and $\hat{x}_i^1$ are ranked in the top two positions.
    - For every $2 \leq j \leq t$, let $\bar{V}_i^{j,\neg}$ be an extension of $O_i^{j,\neg}$ in which $\neg \hat{x}_i^{j-1}$ and $\neg d_i^j$ are ranked in the top two positions.
    - For every $2 \leq j \leq t$, let $\bar{V}_i^j$ be an extension of $O_i^j$ in which $x_i^j$ and $\hat{x}_i^j$ are ranked in the top two positions.

- For every $j \leq t$, if $C_j$ is satisfied by $\mathbf{x}_i = true$ (respectively, $\mathbf{x}_i = false$) for some $i \leq q$, then, we let $\bar{W}_j$ be an extension of $Q_j$ in which $x_i^j$ (respectively, $\neg x_i^j$) is ranked in the second position.

- Let $P^* = \{\bar{V}_1, \ldots, \bar{V}_q\} \cup \{\bar{V}_i^j, \bar{V}_i^{j,\neg} : \forall i \leq q, j \leq t\} \cup \{\bar{W}_1, \ldots, \bar{W}_t\} \cup P_2$.

It can be checked that in $P^* \setminus P_2$, every alternative $c'$ ($c' \neq c$) is ranked in the two top positions at most once. We recall that $s^2(P_2, c') \leq q+t-2$ and $s^2(P^*, c) = q+t$. Therefore, $c$ is the unique winner.

Next, we show how to convert a feasible solution to PW to a feasible solution to the 3-SAT instance. Let $P^*$ be an extension for which $c$ is the unique winner. Let $g$ be the valuation such that for any $i \leq q$, $g(\mathbf{x}_i) = true$ if and only if in the extension of $O_i$ in $P^*$, $\neg x_i$ is ranked in the second position. We prove the following claim to show that under $g$, all clauses are satisfied.

**Claim 1** *For any $i \leq q$, if $g(\mathbf{x}_i) = true$ (respectively, $g(\mathbf{x}_i) = false$), then for every $j \leq t$, $\neg x_i^j$ and $\neg \hat{x}_i^j$ (respectively, $x_i^j$ and $\hat{x}_i^j$) are ranked in the top two positions in the extension of $O_i^{j,\neg}$ (respectively, $O_i^j$) in $P^*$.*

*Proof.* For any $i \leq q$, we prove the claim by induction on $j$. We only prove the case where $g(\mathbf{x}_i) = true$; the case where $g(\mathbf{x}_i) = false$ can be proved similarly.

Suppose $g(\mathbf{x}_i) = true$. By definition, $\neg x_i$ is ranked in the second position in the extension of $O_i$ in $P^*$. We recall that $s^2(P_2, \neg x_i) = q + t - 2 = s^2(P^*, c) - 2$. Because $c$ is the unique winner, $\neg x_i$ is not ranked in the top two positions in any extension of $P_1 \setminus \{O_i\}$. Specifically, $\neg x_i$ is not ranked in the top two positions in the extension of $O_i^{1,\neg}$. We recall that $\neg x_i \succ \neg d_i^1$ in $O_i^{1,\neg}$. Therefore, $\neg d_i^1$ is not ranked in the top two positions in the extension of $O_i^{1,\neg}$ (otherwise, $\neg x_i$ would also be ranked in the top two positions, which immediately prevents $c$ from being the unique winner). We also note that $\neg x_i, \neg d_i^1, \neg x_i^1, \neg \hat{x}_i^1$ are the only four alternatives that can be ranked in the top two positions in an extension of $O_i^{1,\neg}$. It follows that in the extension of $O_i^{1,\neg}$, $\neg x_i^1, \neg \hat{x}_i^1$ are ranked in the top two positions. This means that the claim holds for $j = 1$.

Suppose the claim holds for all $j$ with $j \leq j'$. Following similar reasoning as in the case where $j = 1$, we can prove that the claim holds for $j = j' + 1$. More precisely, by the induction hypothesis, $\neg \hat{x}_i^{j'}$ is ranked in the top two positions in the extension of $O_i^{j',\neg}$. Therefore, $\neg \hat{x}_i^{j'}$ is not ranked in the top two positions in the extension of $O_i^{j'+1,\neg}$ (otherwise the score of $\neg \hat{x}_i^{j'}$ is at least as large as the score of $c$, which means that $c$ is not a unique winner). We recall that $\neg \hat{x}_i^{j'} \succ \neg d_i^{j'+1}$ in $O_i^{j'+1,\neg}$. Therefore, $\neg d_i^{j'+1}$ is not ranked in the top two positions in the extension of $O_i^{j'+1,\neg}$ (otherwise $\neg \hat{x}_i^{j'}$ must also be ranked in the top two positions, which immediately prevents $c$ from being the unique winner). We also note that $\neg \hat{x}_i^{j'}, \neg d_i^{j'+1}, \neg x_i^{j'+1}, \neg \hat{x}_i^{j'+1}$ are the only four alternatives that can be ranked in the top two positions in an extension of $O_i^{j'+1,\neg}$. It follows that in the extension of $O_i^{j'+1,\neg}$, $\neg x_i^{j'+1}$ and $\neg \hat{x}_i^{j'+1}$ are ranked in the top two positions. This means that the claim holds for $j = j' + 1$.

Therefore, the claim holds for every $j \leq t$. □

We are now ready to show that under $g$, all the clauses are satisfied. Let $j$ be a number no more than $t$. If $x_i^j$ is ranked in the second position in the extension of $Q_j$, then we must have that $g(\mathbf{x}_i) = true$. If not, then, from Claim 1, $x_i^j$ is ranked in the top two positions in the extension of $O_i^j$, which means that $x_i^j$ is ranked in the top two positions in $P^* \setminus P_2$ at least twice: once in $O_i^j$, and once in $Q_j$. It follows that $s^2(P^*, x_i^j) \geq q+t-2+2 \geq q+t = s^2(P^*, c)$, which contradicts the assumption that $c$ is the unique winner. Similarly, if in the extension of $Q_j$, $\neg x_i^j$ is ranked in the second position, then we must have that $g(\mathbf{x}_i) = false$. This means that under $g$, every clause $C_j$ is satisfied by the valuation of the variable that corresponds to the alternative that is ranked in the second position in the extension of $Q_j$. Hence, $F$ is satisfied.

For PcW, we simply replace $s^2(P_2, x_i) = s^2(P_2, \neg x_i) = s^2(P_2, x_i^j) = s^2(P_2, \neg x_i^j) = s^2(P_2, \hat{x}_i^j) = s^2(P_2, \neg \hat{x}_i^j) = q + t - 2$ in the definition for $P_2$ by $s^2(P_2, x_i) = s^2(P_2, \neg x_i) = s^2(P_2, x_i^j) = s^2(P_2, \neg x_i^j) = s^2(P_2, \hat{x}_i^j) = s^2(P_2, \neg \hat{x}_i^j) = q + t - 1$.

The reduction for $k > 2$ is similar to the case where $k = 2$. For any 3-SAT instance, let $P_1$ and $P_2$ be the profile of partial orders defined for the case $k = 2$. For $k > 2$, we add $|P_1 \cup P_2| \times (k - 2)$ new alternatives to the instance, and in each partial order in $P_1 \cup P_2$, we let the top $k - 2$ positions be occupied by the new alternatives, and we put the remaining new alternatives in the bottom positions, such that none of the new alternatives is ranked in the top $k$ positions more than once. Let $\bar{P}_1$ and $\bar{P}_2$ denote the profiles of partial orders obtained in this way. It follows that $c$ is a possible (co-)winner for $\bar{P}_1 \cup \bar{P}_2$ with respect to $k$-approval if and only if $c$ is a possible (co-)winner for $P_1 \cup P_2$ with respect to 2-approval. $\square$

**Theorem 3** *PW and PcW are NP-complete and NW and NcW are coNP-complete with respect to Copeland, even when the number of undetermined pairs in each vote is at most 8.*

*Proof.* We first prove the PW and NcW parts, in one reduction from X3C. Without loss of generality, we can always assume that in the X3C instance, $t$ is odd and $t = q$, because if not, then we make the following changes to the X3C instance.

- If $t > q$, then we add $3(t - q)$ dummy elements $v'_1, \ldots, v'_{3(t-q)}$ and $2(t - q)$ sets $S'_1, S'_1, \ldots, S'_{t-q}, S'_{t-q}$, where for each $i \leq t - q$, $S'_i = \{v'_{3i-2}, v'_{3i-1}, v'_{3i}\}$.

- If $q > t$, then we add $q - t$ copies of $S_1$.

- If $q = t$ and $t$ is even, then we add three dummy elements $v'_1, v'_2, v'_3$, and three copies of $S'_1 = \{v'_1, v'_2, v'_3\}$.

In the new X3C instance, $t = q$, $t$ is odd, the size of the instance is polynomial in the size of the old one, and the new X3C instance has a feasible solution if and only if the old one has.

Given an X3C instance $\mathcal{V} = \{v_1, \ldots, v_q\}$, $\mathcal{S} = \{S_1, \ldots, S_t\}$, where $q = t$ and $t$ is odd, we construct a PW instance as follows.

**Alternatives:** $\{c, w, d\} \cup \mathcal{V} \cup A \cup B$, where $A = \{a_1, \ldots, a_{t-2}\}$, $B = \{b_1, \ldots, b_{7t}\}$.

**First part $P_1$ of the profile:** Let $M$ be a cyclic permutation among $B$. That is, $M = b_1 \to b_2 \to \ldots \to b_{7t} \to b_1$. Let $V_B = b_1 \succ b_2 \succ \ldots \succ b_{7t}$. For each $i \leq t$, we obtain a partial order by starting with $O((\mathcal{V} \setminus S_i), d, S_i, w, c, M^i(V_B), A)$, and then removing the ordering relationships in $(\{d\} \cup S_i) \times \{w, c\}$.

**Second part $P_2$ of the profile:**

- $t - \dfrac{2q}{3} + 1$ votes: for each $i$ such that $t + 1 \leq i \leq 2t - \dfrac{2q}{3} + 1$, there is a vote that is consistent with $w \succ c \succ d \succ \mathcal{V} \succ M^i(V_B) \succ A$.

- $\frac{q}{3} - 2$ votes: for each $i$ such that $2t - \frac{2q}{3} + 2 \leq i \leq 2t - \frac{q}{3} - 1$, there is a vote that is consistent with $w \succ c \succ d \succ \mathcal{V} \succ M^i(V_B) \succ A$.

- $\frac{q}{3} - 2$ votes: for each $i$ such that $2t - \frac{q}{3} \leq i \leq 2t - 3$, there is a vote that is consistent with $w \succ d \succ c \succ \mathcal{V} \succ M^i(V_B) \succ A$.

- 2 votes: for each $i$ such that $2t - 2 \leq i \leq 2t - 1$, there is a vote that is consistent with $c \succ w \succ d \succ \mathcal{V} \succ M^i(V_B) \succ A$.

- 2 votes: for each $i$ such that $2t \leq i \leq 2t + 1$, there is a vote that is consistent with $d \succ c \succ \mathcal{V} \succ w \succ M^i(V_B) \succ A$.

- $\frac{1}{2}(5t - 1)$ votes: for each $i$ such that $2t + 2 \leq i \leq \frac{1}{2}(9t + 1)$, there is a vote that is consistent with $w \succ A \succ c \succ M^i(V_B) \succ \mathcal{V} \succ d$.

- $\frac{1}{2}(5t - 1)$ votes: for each $i$ such that $\frac{1}{2}(9t + 3) \leq i \leq 7t$, there is a vote that is consistent with $M^i(V_B) \succ \mathcal{V} \succ w \succ d \succ A \succ c$.

We note that the number of undetermined pairs in each vote is no more than 8.

Let $P_1'$ denote the profile that extends $P_1$ such that in each vote $d$ and $S_i$ are ranked higher than $w$ and $c$, that is, $P_1' = \{O((\mathcal{V} \setminus S_i), d, S_i, w, c, B, A) : i \leq t\}$. We make the following observations on each pairwise election:

- $w$ always defeats $c, d, B, A$, and for each $i \leq q$, $D_{P_1' \cup P_2}(v_i, w) = 3$.

- $c$ always defeats $\mathcal{V}, B$, always loses to $A$, and $D_{P_1' \cup P_2}(d, c) = \frac{2q}{3} - 1$.

- $B$ always defeats $d, \mathcal{V}, A$, and due to its cyclic order in the profile, $b_j$ always defeats $b_{j+1}, \ldots, b_{j+\frac{1}{2}(7t-1)}$, where for any $i \in \mathbb{N}$, $b_i = b_{i+7t}$, and always loses to the other alternatives in $B$.

Therefore, in $P_1' \cup P_2$, the total number of pairwise elections won by each alternative is:

- $w$ wins $|B| + |A| + 2 = 8t$,

- $c$ wins $|\mathcal{V}| + |B| = q + 7t = 8t$,

- $d$, any $v \in \mathcal{V}$, and any $a \in A$ wins at most $8t + q + 1 - 7t = t + q + 1$, because they all lose to $B$,

- any $b \in B$ wins at most $\frac{1}{2}(|B| - 1) + |A| + |\mathcal{V}| + 1 = \frac{1}{2}(9t + 2q - 3)$ pairwise elections.

We recall that in the X3C instance $t = q$, which means in $P_1' \cup P_2$, the winners are $\{w, c\}$. In order for $c$ to be the unique winner, the only possibility is for $c$ to win the pairwise election against $d$ by putting $c \succ d$ in at least $\frac{q}{3}$ votes in $P_1$. However, when we put $c$ ahead of $d$ in a vote corresponding to $S_i$, for all $v \in S_i$ the pairwise score difference between $w$ and $v$ increases by 2. Moreover, if $w \succ v$ for some $v \in \mathcal{V}$ at least twice in an extension $P^*$ of $P_1$, then $D_{P^* \cup P_2}(v, w) \leq -1$, which means that $w$ defeats $v$ in their pairwise election. In this case, $w$ would win $8m + 1$ pairwise elections, which means that $c$ cannot be the unique winner. Therefore, $c$ is a possible unique winner if and only if there exists an extension $P^*$ of $P_1$ such that $c \succ d$ in exactly $\frac{q}{3}$ votes in $P^*$, and the corresponding $S_i$ do not overlap, that is, they constitute an exact cover of $\mathcal{V}$. This means that PW has a solution if and only if the X3C problem has a solution. So PW is NP-complete.

Because in the above reduction, $w$ would always be a co-winner if $c$ is not the unique winner, NcW is coNP-complete. For PcW and NW, we just need to slightly modify the reduction for PW and NcW: let $|A| = t - 1$ and keep the rest unchanged. Then, $w$ will initially win $8t + 1$ pairwise elections, and $c$ is a possible co-winner ($w$ is not the necessary unique winner) if and only if there exists a feasible solution to the X3C instance. □

**Theorem 4** *PW and PcW are NP-complete with respect to Bucklin, even when the number of undetermined pairs in each vote is at most 16.*

*Proof.* First, we give a reduction from X3C to PW. Given any X3C instance $\mathcal{V} = \{v_1, \ldots, v_q\}$, $\mathcal{S} = \{S_1, \ldots, S_t\}$, we construct a PW instance as follows.

**Alternatives:** $W \cup D \cup \mathcal{V} \cup \{c, w\}$, where $W = \{w_1, \ldots, w_{q+1}\}$, $D = \{d_1, \ldots, d_{q+1}\}$.

**First part $P_1$ of the profile:** for each $i \leq t$, we start with $O(w_1, \ldots, w_{q+1}, S_i, c, (\mathcal{V} \setminus S_i), D)$, and then obtain a partial order by removing the relations in

$$\{w_{q-2}, w_{q-1}, w_q, w_{q+1}\} \times (S_i \cup \{c\})$$

**Second part $P_2$ of the profile:**

- $t$ copies of $\mathcal{V} \succ c \succ$ Others,
- $\frac{q}{3} - 1$ copies of $\mathcal{V} \succ w \succ c \succ$ Others,
- $\frac{q}{3} + 2$ copies of $D \succ w_1 \succ$ Others.

We note that the number of undetermined pairs in each vote is no more than 16. Notice $|P_1 \cup P_2| = 2t + \frac{2q}{3} + 1$, and $w_1$ is ranked within top $q + 2$ positions in $t + \frac{q}{3} + 2$ votes in any extension of $P_1 \cup P_2$. Therefore, in order for $c$ to win, $c \succ w_{q-2}$ must hold in at least $\frac{q}{3}$ votes in the extension of $P_1$. However, whenever we put $c$ ahead of $w_{q-2}$ in a vote, we are forcing the alternatives in the $S_i$ corresponding to that vote be ranked within top $q$ positions. If some $v \in \mathcal{V}$ is ranked within top $q$ positions at least twice in an extension of $P_1$, then overall it will be ranked within top $q$ positions in at least $t + \frac{q}{3} + 1$ votes, which means $c$ will not be the unique winner.

If there exists a feasible solution to the X3C problem, then we can put $c$ ahead of $w_{q-2}$ in the votes corresponding to this solution, so that we obtain an extension $P_1^*$ of $P_1$ such that $c$ is ranked within top $q + 1$ positions in $\frac{q}{3}$ votes, while for any $v \in \mathcal{V}$, $v$ is ranked within top $q$ (and, in fact, the first $q + 1$) positions just once. As a result, $c$ is the unique winner of the profile $P_1^* \cup P_2$, because no other alternative is ranked within top $q + 1$ positions in at least $t + \frac{q}{3}$ votes. Conversely, if $c$ is the unique winner in some profile $P_1^* \cup P_2$, then $P_1^*$ corresponds to a feasible solution to the X3C problem. Therefore, PW with respect to Bucklin is NP-complete.

For PcW, we just need to modify the reduction slightly, by changing the last $\frac{q}{3}+1$ votes from $[D \succ w_1 \succ \text{Others}]$ to $[d_1 \succ \ldots \succ d_q \succ w_1 \succ \text{Others}]$. In this case, the Bucklin score of $w_1$ is $q+1$, which means $c$ can at best hope to be a co-winner. As a result, PcW is also NP-complete. □

To prove our hardness results for maximin, ranked pairs, and voting trees, we present two helpful lemmas. We first show that given any pair of alternatives $c, c'$, there exist two linear orders that increase $D(c, c')$ by two while keeping all other pairwise score differences unchanged. This lemma has been used previously (McGarvey, 1953; Conitzer & Sandholm, 2005a). We will use this technique in the second (score-adjusting) part of the reductions for maximin, ranked pairs, and voting trees.

**Lemma 1** *Given any profile $P$ and any pair of different alternatives $c, c'$, let the remaining alternatives be $\{c_1, \ldots, c_{m-2}\}$. Let $P'$ be the profile consisting of $P$ plus the following two votes:*

1. $[c \succ c' \succ c_1 \succ \ldots \succ c_{m-2}]$, and

2. $[c_{m-2} \succ \ldots \succ c_1 \succ c \succ c']$.

*Then, $D_{P'}(c, c') = D_P(c, c') + 2$, and for any alternatives $d, d'$ such that $\{d, d'\} \neq \{c, c'\}$, $D_{P'}(d, d') = D_P(d, d')$.*

This lemma tells us that the pairwise score differences can be changed almost arbitrarily. The only constraint is that the parity of the pairwise score differences remains the same. The following lemma is a direct corollary.

**Lemma 2** *(The main theorem in McGarvey, 1953) Given a profile $P$ and any skew-symmetric function $F : \mathcal{C} \times \mathcal{C} \to \mathbb{Z}$ (that is, $F(c_1, c_2) = -F(c_2, c_1)$ for all $c_1, c_2$), such that for all pairs of alternatives $c, c' \in \mathcal{C}$, $F(c, c') - D_P(c, c')$ are all even (or all odd), then there exists a profile $P'$ such that*

1. $|P'| \leq \frac{1}{2} \sum_{c,c'} (|F(c, c') - D_P(c, c')| + 1)$,

2. $D_{P \cup P'} = F$.

That is, for any skew-symmetric function $F$ such that for all pairs of alternatives $(c, c')$ (with $c \neq c'$), $F(c, c') - D_P(c, c')$ has the same parity, we can change the pairwise score differences from $D_P$ to $F$ by adding no more than $\frac{1}{2} \sum_{c,c'} (|F(c, c') - D_P(c, c')| + 1)$ votes to $P$. Here, the factor $\frac{1}{2}$ comes from the fact that for any pair of alternatives $c$ and $c'$, the absolute value of the difference between $F$ and $D_P$ is counted twice, i.e., $|F(c, c') - D_P(c, c')| = |F(c', c) - D_P(c', c)|$. In fact, it is possible to obtain even tighter bounds on the needed size of $P'$ (Erdös & Moser, 1964), but for the purpose of our NP-hardness proofs this does not matter.

Now we are ready to prove the hardness results for maximin and ranked pairs. As we mentioned in the beginning of this section, in all hardness proofs in this section, the profile consists of $P_1$ and $P_2$, where $P_1$ is a set of partial orders used to encode the X3C instance, and $P_2$ is a set of linear orders used to adjust the "scores" of the alternatives. For maximin, ranked pairs, and voting trees, $P_2$ is used to adjust the pairwise score differences. We do not explicitly give $P_2$ in the reductions for these rules. Instead, we present the properties of $P_2$, then appeal to Lemma 2 to assert that $P_2$ does exist, and can be constructed in polynomial time.

**Theorem 5** *PW and PcW are NP-complete with respect to maximin, even when the number of undetermined pairs in each vote is at most 4.*

*Proof.* We first prove that PW is NP-complete. Given an X3C instance $\mathcal{V} = \{v_1, \ldots, v_q\}$, $\mathcal{S} = \{S_1, \ldots, S_t\}$, we construct a PW instance as follows.

**Alternatives:** $\mathcal{V} \cup \{c, w, w'\}$.

**First part $P_1$ of the profile:** for each $i \leq t$, we start with $O(w, S_i, c, (\mathcal{V} \setminus S_i), w')$, and subsequently obtain a partial order $O_i$ by removing the relations in $\{w\} \times (S_i \cup \{c\})$.

**Second part $P_2$ of the profile:** according to Lemma 2, $P_2$ is defined to be a set of votes such that the pairwise score differences of $\{O(w, S_i, c, (\mathcal{V} \setminus S_i), w') : i \leq t\} \cup P_2$ satisfy:

(1) $D(w, c) = t + \frac{2q}{3} - 2$; for each $i \leq q$, $D(w, v_i) = t + 2$; $D(w', w) = D(v_1, w') = t + 4$; $D(w', c) = t - 2$.

(2) For all other pairwise scores not defined in (1), $D(l, r) \leq 1$.

We note that the number of undetermined pairs in each vote is no more than 4. Lemma 2 implies that the size of $P_2$ is polynomial in $q + t$.

We note that the minimum pairwise score difference of $w$ is $D(w, w') = -t - 4$; the minimum pairwise score difference of $w'$ is also $-t - 4 = D(w', v_1)$.

Suppose there exists a profile $P_1^*$ extending $P_1$ such that $c$ wins in $P_1^* \cup P_2$. If $c$ is raised higher than $w$ in at least one and at most $\frac{q}{3} - 1$ votes in $P_1^*$, then, $D(c, w) \leq -t$, and there exists $i \leq q$ such that $D(v_i, w) \geq -t$ (the smallest pairwise score difference of $v_i$), which means that $c$ is not the unique winner because $v_i$ is performing at least as well. If $c$ is ranked higher than $w$ in at least $\frac{q}{3} + 1$ votes in $P_1^*$, then we still have $D(c, w') = -t + 2$, and there exists $i \leq q$ such that $v_i$ is ranked higher than $w$ in at least two votes in $P_1^*$, which means that $D(v_i, w) \geq -t + 2$ (the smallest pairwise score difference of $v_i$). It follows that in this case, $c$ is not the unique winner because $v_i$ is performing at least as well. Therefore, the only way for $c$ to win is to decrease $D(w, c)$ by raising $c$ higher than $w$ in exactly $\frac{q}{3}$ votes in $P_1^*$. However, each time that we decrease $D(w, c)$ by 2 due to adding $c \succ w$ to $O_i \in P_1$, for each $v \in S_i$, $D(w, v)$ is also decreased by two. Because $D(w', c) = t - 2$, decreasing $D(w, c)$ to less than $t - 2$ would not raise the minimum pairwise score difference of $c$. But if for some $i \leq q$, $D(w, v_i)$ is decreased by 4 or more, then the minimum pairwise score of $v_j$ is at least $-t+2$, which means that in this case $c$ cannot be the unique winner. Therefore, the sets $S_i$ in the votes in $P_1^*$ where $c \succ w$ cannot overlap. Because there must be at least $q/3$ of these votes, the corresponding subsets $S_i$ constitute a feasible solution to the X3C instance.

Conversely, suppose the X3C instance has a solution. Without loss of generality, let the solution be $\{S_1, \ldots, S_{q/3}\}$. We define an extension $P_1^*$ of $P_1$ by adding $c \succ w$ in $O_i$ for all $i \leq q/3$, and then adding $w \succ S_i$ for all $i > q/3$. It follows that $c$ is the unique winner for the profile $P_1^* \cup P_2$ with respect to the maximin rule. Therefore PW is NP-complete.

For PcW, we just need to slightly modify the above reduction: we replace the condition $D(w, v_i) = t + 2$ by $D(w, v_i) = t$ when constructing $P_2$. Therefore PcW is NP-complete. □

**Theorem 6** *PW and PcW are NP-complete and NW and NcW are coNP-complete with respect to ranked pairs, even when the number of undetermined pairs in each vote is at most 8.*

*Proof.* We first prove the NP-hardness of PW and NcW in one reduction. Given an X3C instance $\mathcal{V} = \{v_1, \ldots, v_q\}$, $\mathcal{S} = \{S_1, \ldots, S_t\}$, we construct a PW instance as follows.

**Alternatives:** $\mathcal{V} \cup \{c, a, b, w\}$.

**First part $P_1$ of the profile:** for each $i \leq t$, we start with $O(a, c, S_i, b, \text{Others})$, and subsequently obtain a partial order $O_i$ by removing the relations in $(\{a, c\} \times (S_i \cup \{b\}))$.

**Second part $P_2$ of the profile:** according to Lemma 2, $P_2$ is defined to be a set of votes such that the pairwise score differences of $\{O(a, c, S_i, b, \text{Others}) : i \leq t\} \cup P_2$ satisfy:

1. For all $i \leq q$, $D(c, b) = D(w, a) = D(w, v_i) = 3t + \frac{2q}{3}$.
2. $D(a, c) = t + \frac{2q}{3}$, $D(c, w) = t + \frac{2q}{3} - 2$, $D(v_i, c) = t + \frac{2q}{3} - 6$, $D(b, a) = t + 2$.
3. $D(l, r) = 0$ in all other cases.

We note that the number of undetermined pairs in each vote is no more than 8. Lemma 2 implies that the size of $P_2$ is polynomial in $q + t$.

We note that $D(c, b)$, $D(w, a)$, and $D(w, v_i)$ (for every $i \leq q$) are much larger than the remaining pairwise score differences in any extension of $P_1 \cup P_2$. Therefore, $c \succ b$, $w \succ a$, and $w \succ v_i$ (for every $i \leq q$) are fixed first in any extension of $P_1 \cup P_2$. It follows that in the output (a linear order over $\mathcal{C}$) for any extension of $P_1 \cup P_2$, we must have that $c \succ b$, $w \succ a$, and $w \succ v_i$ (for every $i \leq q$). We note that the only way for $c$ to be the unique winner is to lock $b \succ a$ before $a \succ c$. That is, $D(b, a)$ must be at least $t + 2 + \frac{2q}{3}$. However, whenever we let $b \succ a$ in an extension of $O_i$, we are forcing $S_i \succ c$. Let $P_1^*$ be an extension of $P_1$ such that $c$ is the unique winner for the profile $P_1^* \cup P_2$ (or, equivalently, such that $w$ is not a co-winner for the profile $P_1^* \cup P_2$). We note that if there exists $i \leq q$ such that $v_i \succ c$ in at least two votes in $P_1^*$, then $D(v_i, c) \geq t + \frac{2q}{3} - 6 + 4 = t + \frac{2q}{3} - 2 = D(c, w)$, which means that $w$ is a co-winner (by locking $v_i \succ c$ before $c \succ w$). Therefore, in $P_1^*$, we must have that $b \succ a$ in exactly $\frac{q}{3}$ votes, and for all $i \leq q$, $v_i \succ c$ in exactly one vote. This naturally corresponds to a solution to the X3C instance.

Conversely, suppose that the X3C instance has a solution. Without loss of generality, let the solution be $\{S_1, \ldots, S_{q/3}\}$. We define an extension $P_1^*$ of $P_1$ by adding $b \succ a$ in $O_i$ for all $i \leq q/3$, and then for all $i > q/3$, letting the extension of $O_i$ be $[a \succ c \succ S_i \succ b \succ \text{Others}]$. It follows that $c$ is the unique winner for this profile (and hence, $w$ is not a co-winner). Therefore, PW is NP-complete and NcW is coNP-complete with respect to ranked pairs.

For PcW and NW, we just need to slightly modify the above reduction by letting $D(b,a) = t$ and for all $i \leq q$, letting $D(v_i, c) = t + \frac{2q}{3} - 4$. □

Next, we consider voting trees. Because a voting tree is defined for a fixed number of alternatives, to study the complexity of the possible/necessary winner problems with respect to voting trees, we need to consider an infinite sequence of trees, one for each natural number (representing the number of alternatives).[8] Therefore, we let a voting tree rule $T$ be composed of an infinite sequence of voting trees $\{T_1, T_2, \ldots\}$, where for any $m \in \mathbb{N}$, $T_m$ is a voting tree for $m$ alternatives (that is, $T_m$ is a binary tree that has $m$ leaf nodes, and each leaf is associated with an alternative).

For any $t \in \mathbb{N}$, a voting tree $T_m$ is $t$-well-spread if there exist $t$ pairs of leaves $(c_1, a_1), \ldots, (c_t, a_t)$, such that for each $i \leq t$, $c_i$ and $a_i$ are siblings. We say that any leaf in such a pair is a *rich leaf*. A voting tree is *balanced* if the depths of any pair of leaves differ at most by one, and the number of leaves whose (unique) sibling is not a leaf is at most one.

**Example 2** *Two voting trees are illustrated in Figure 2. The voting tree in (a) is 1-well-spread, and $c_1$ and $c_2$ are rich leaves; the voting tree in (b) is balanced and 3-well-spread, and all leaves except $c_5$ are rich leaves.*

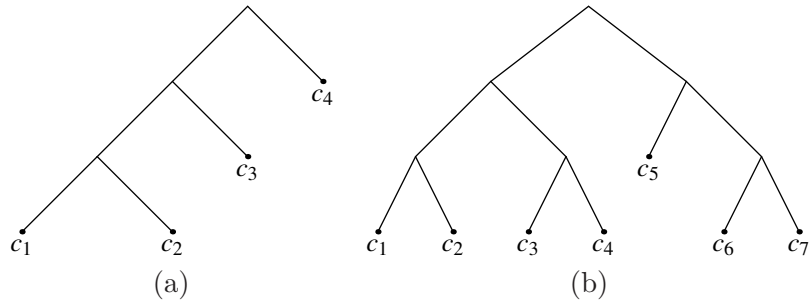

Figure 2: Voting trees.

**Theorem 7** *For any voting tree rule $T = \{T_1, T_2, \ldots\}$, if there exists a polynomial function $f(x)$ such that for any $x \in \mathbb{N}$, there exists $l \in \mathbb{N}$ with $x \leq l \leq f(x)$ such that $T_l$ is $x$-well-spread, then PW and PcW are NP-complete, and NW and NcW are coNP-complete with respect to $T$, even when the number of undetermined pairs in each vote is at most 16.*

---

8. This is similar to the case of positional scoring rules, which are technically defined only for a specific number of alternatives.

*Proof.* Let $j_2, j_3, \ldots$ be the index of the voting trees such that for any $z \in \mathbb{N}$ ($z \geq 2$), $T_{j_z}$ is $2(z+1)$-well-spread and $j_z \leq f(2(z+1))$. For any $z$, we let $c$ be an arbitrary rich leaf in $T_{j_z}$.

We first prove the NP-hardness of PW and PcW in a single reduction. Given an X3C instance $\mathcal{V} = \{v_1, \ldots, v_q\}$, $\mathcal{S} = \{S_1, \ldots, S_t\}$, we construct a PW instance as follows.

**Alternatives:** Let $\mathcal{C}$ be the leaves of $T_{j_q}$, where $\mathcal{C} = \{c, d, w\} \cup \mathcal{V} \cup A \cup E$, and $A = \{a_1, \ldots, a_q\}$, $E = \{e_1, \ldots, e_{m_q - 2q - 3}\}$, where $m_q$ is the number of leaves in $T_{j_q}$. Let the tree be such that $\{c, d\} \cup \mathcal{V} \cup A$ are rich leaves in a subtree whose root is a child of the root of $T_{j_q}$ (because $T_{j_q}$ is $2(q+1)$-well-spread, this is always possible); $d$ is the sibling of $c$; the only common ancestor of $c$ and $w$ is the root; and for each $1 \leq i \leq q$, $v_i$ and $a_i$ are siblings. The positions of $\{c, d, w\} \cup \mathcal{V} \cup A$ are illustrated in Figure 3. $E$ is the set of all other alternatives in $T_{j_q}$. For each $i \leq t$, if $S_i = \{v_{l(i,1)}, v_{l(i,2)}, v_{l(i,3)}\}$, then we let $A_i = \{a_{l(i,1)}, a_{l(i,2)}, a_{l(i,3)}\}$—that is, $A_i$ consists of the siblings of the elements in $S_i$.

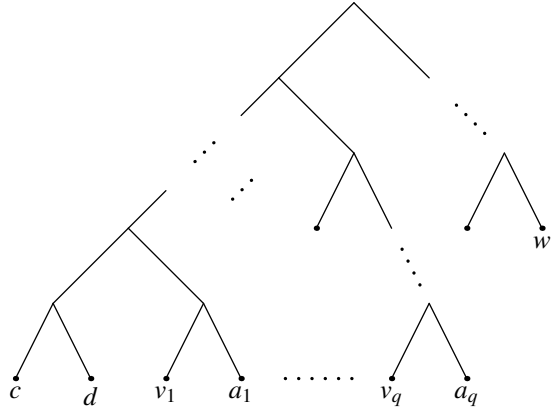

Figure 3: Positions of the alternatives in $T_{j_q}$.

**First part $P_1$ of the profile:** for each $i \leq t$, we start with $O(d, A_i, S_i, c, \text{Others})$, and subsequently obtain a partial order $O_i$ by removing relations in $(\{d\} \cup A_i) \times (S_i \cup \{c\})$.

**Second part $P_2$ of the profile:** according to Lemma 2, $P_2$ is defined to be a set of votes (linear orders) such that the pairwise score differences for the profile $\{O(d, A_i, S_i, c, \text{Others}) : i \leq t\} \cup P_2$ satisfy:

(1) $D(c, d) = -2q/3 + 1$, $D(c, w) = 2q + 1$.

(2) For each $i \leq q$, $D(a_i, v_i) = 3$, $D(v_i, c) = D(c, a_i) = 2q + 1$.

(3) For each $c' \in \mathcal{C}$ (with $c' \neq c$), $D(w, c') = 2q + 1$.

(4) For each pair $i, i' \leq q$ (with $i \neq i'$), $D(v_i, a_{i'}) = 2q + 1$.

(5) For each $x \in \mathcal{C} \setminus E$, and each $e \in E$, $D(x, e) = 2q + 1$.

We note that the number of undetermined pairs in each vote is no more than 16. Lemma 2 implies that the size of $P_2$ is polynomial in $q + t$.

The only way for $c$ to win is to beat $d$ in the first round, and not to meet any of $\{v_1, \ldots, v_q\}$ in later rounds, which can only happen if every $v_i$ is beaten by the corresponding $a_i$ in the first round. This is because by item (4), for each $i \neq i'$, $D(v_i, a_{i'}) = 2q + 1$, which means that if for some $i \leq q$, $v_i$ wins in the first round, it will only be beaten by $w$ or $v_j$ for some $j \leq q$ in subsequent rounds. In this case the winner must be $w$. It follows that in any extension of $P_1$ that makes $c$ win, $c$ must be ranked higher than $d$ at least $q/3$ times. However, if we rank $c$ higher than $d$ in an extension of $O_i$, then in the same extension we must have that $S_i \succ A_i$. In order for every $a_i$ to defeat $v_i$, for every $i \leq q$, $v_i$ can be ranked higher than $a_i$ at most once in the extension of $P_1$. Therefore, if there exists a profile $P_1^*$ extending $P_1$ such that $c$ is the unique winner (or co-winner) in $P_1^* \cup P_2$, then the votes in $P_1^*$ where $c \succ d$ make up a feasible solution to the X3C problem instance. Conversely, for any feasible solution to the X3C problem instance, we can find a $P_1^*$ extending $P_1$ such that $c$ is the unique winner of the profile $P_1^* \cup P_2$ with respect to $T_{i_j}$. Therefore, PW and PcW are NP-complete.

Because if $c$ is not the unique winner, then $w$ is always the unique winner. Therefore, NW and NcW are coNP-complete. □

From Theorem 7, we immediately obtain the following hardness results for voting tree rules composed of balanced trees, by setting $f(x) = 4x$ (because there will exist some integer $y$ such that $2x \leq 2^y \leq 4x$, so in the balanced tree for $2^y$ alternatives there will be at least $x$ pairs of siblings).

**Corollary 2** *PW and PcW are NP-complete and NW and NcW are coNP-complete with respect to the voting tree rule that is composed of balanced binary trees, even when the number of undetermined pairs in each vote is at most 16.*

Finally, we have the following theorems on the complexity of PW and NcW with respect to plurality with runoff.

**Theorem 8** *PW is NP-complete with respect to plurality with runoff.*

*Proof.* We prove NP-hardness by a reduction from X3C. Given an X3C instance $\mathcal{V} = \{v_1, \ldots, v_q\}$, $\mathcal{S} = \{S_1, \ldots, S_t\}$, we construct a PW instance as follows.

**Alternatives:** $\mathcal{C} = \{c, d, e\} \cup \mathcal{S}_V \cup E$, where $\mathcal{S}_V = \{s_1, \ldots, s_t\}$ and $E = \{e_1, \ldots, e_{(q+4)^2(t+4)^4}\}$.

**First part $P_1$ of the profile:** $P_1 = P_1^1 \cup P_1^2$, where $P_1^1$ and $P_1^2$ are defined as follows.

- $P_1^1$: for each $i \leq q$, we start with a linear order $O(d, \mathcal{S}_V, c, \text{Others})$, and subsequently obtain a partial order $O_i$ by removing $(\{d\} \cup S_V) \times \{s_j : v_i \in S_j\}$. That is, we remove a minimum set of constraints such that any alternative in $\{s_j : v_i \in S_j\}$ can be ranked in the top position in at least one extension of $O_i$. Let $P_1^1 = \{O_i : i \leq q\}$.
- $P_1^2$: for each $j \leq t$, we start with a linear order $O(d, e, c, \text{Others})$, and subsequently obtain a partial order $Q_j^1$ by removing $(\{d\} \times \{e\}) \cup (\mathcal{C} \times \{s_j\})$. That is, in an extension of $Q_j^1$, only $d, e$, and $s_j$ can be ranked in the top position. We let $Q_j^2 = Q_j^1$, and $P_1^2 = \{Q_j^1 : j \leq t\} \cup \{Q_j^2 : j \leq t\}$.

**Second part $P_2$ of the profile:** $P_2 = P_2^1 \cup P_2^2$, where $P_2^1$ and $P_2^2$ are defined as follows.

- $P_2^1$: a set of $q(t + 7/3) + 8$ votes, in which $c$ is ranked in the top position $q + 4$ times, $d$ is ranked in the top position $q + 2$ times, $e$ is ranked in the top position $q/3 + 2$ times, and for each $j \leq t$, $s_j$ is ranked in the top position $q$ times. It does not matter how the remaining alternatives are ranked in $P_2^1$.

- $P_2^2$: we first obtain, according to Lemma 2, a profile $\hat{P}_2^2$ such that the pairwise score differences of the following profile:

$$\{q \text{ copies of } O(d, \mathcal{S}_V, c, \text{Others})\} \cup \{2t \text{ copies of } O(d, e, c, \text{Others})\} \cup P_2^1 \cup \hat{P}_2^2$$

  satisfy the following conditions.
  1. $D(d, c) = D(e, c) = 1$;
  2. for all $j \leq t$, $D(c, s_j) = 1$.

  By Lemma 2, the size of $\hat{P}_2^2$ is polynomial in $p + t$. Next, we obtain $P_2^2$ from $\hat{P}_2^2$ by moving an alternative in $E$ to the top position in each vote of $\hat{P}_2^2$, in such a way that each vote in $P_2^2$ ranks a different alternative in the top position. $P_2^2$ is well-defined, because $|E| \geq |\hat{P}_2^2|$.

For any profile $P$, and any alternative $c'$, we let $Plu_P(c')$ denote the plurality score of $c'$ in $P$, that is, $Plu_P(c')$ is the number of times where $c'$ is ranked in the top position in $P$. The subscript $P$ is omitted when there is no risk of confusion. We make the following observations on the profile $\{q \text{ copies of } O(d, \mathcal{S}_V, c, \text{Others})\} \cup \{2t \text{ copies of } O(d, e, c, \text{Others})\} \cup P_2^1 \cup P_2^2$:

- $D(d, c) = D(e, c) = 1$, and for all $j \leq t$, $D(c, s_j) = 1$;

- $Plu(c) = q + 4$, $Plu(d) = 2t + 2q + 2$, $Plu(e) = q/3 + 2$; for each $j \leq t$, $Plu(s_j) = q$; for each $e' \in E$, $Plu(e') \leq 1$.

We also note that in any extension of $P_1 \cup P_2$, $Plu(c) = q + 4$.

If the X3C instance has a solution $S_{j_1}, \ldots, S_{j_{q/3}}$, then we construct a solution to the PW instance as follows.

- For each $i \leq q$, let $V_i = [s_{j_l} \succ d \succ (\mathcal{S}_V \setminus \{s_{j_l}\}) \succ c \succ \text{Others}]$, where $j_l$ is such that $c_i \in S_{j_l}$; we note that $V_i$ extends $O_i$;

- for each $l \leq q/3$, let $V_{j_l}^1 = V_{j_l}^2 = [e \succ d \succ c \succ \text{Others}]$; we note that $V_{j_l}^1$ and $V_{j_l}^2$ extend $Q_{j_l}^1$ and $Q_{j_l}^2$, respectively;

- for each $j \leq t$ (with $j \neq j_l$ for all $l \leq q/3$), let $V_j^1 = V_j^2 = [s_j \succ d \succ e \succ c \succ \text{Others}]$; we note that $V_j^1$ and $V_j^2$ extend $Q_j^1$ and $Q_j^2$, respectively;

- then, we use these votes to extend the partial orders in $P_1$: let $P_1^* = \{V_i : i \leq q\} \cup \{V_j^1, V_j^2 : j \leq t\}$.

In $P_1^* \cup P_2$, we have $Plu(c) = q + 4$, $Plu(d) = Plu(e) = q + 2$; for each $l \leq q/3$, $Plu(s_{j_l}) = q + 3$; for each $j \neq j_l$ ($l = 1, \ldots, q/3$), $Plu(s_j) = q + 2$; and for each $e' \in E$, $Plu(e') \leq 1$. Also, we have that for each $l \leq q/3$, $D(c, s_{j_l}) = 1$. It follows that the pairs that enter the runoff (in some parallel universe) are $(c, s_{j_1}), \ldots, (c, s_{j_{q/3}})$, and $c$ wins all of these pairwise elections. Therefore, $c$ is the unique winner for $P_1^* \cup P_2$.

Next, we show how to convert a solution to the PW instance to a solution to the X3C instance. Let $P_1^* = P_1^{1*} \cup P_1^{2*}$ be an extension of $P_1$ such that $c$ is the unique winner for $P_1^* \cup P_2$, where $P_1^{1*} = \{V_i : i \leq q\}$ extends $P_1^1$, and $P_1^{2*} = \{V_j^1 : j \leq t\} \cup \{V_j^2 : j \leq t\}$ extends $P_1^2$. We make the following sequence of claims.

**Claim 2** *Neither $d$ nor $e$ can enter the runoff, which means that the only pairs that could potentially still enter the runoff are $(c, s_j)$, for some $j \leq t$.*

*Proof.* If $d$ or $e$ entered the runoff in some parallel universe, then it would defeat $c$ in the runoff (unless $c$ is not even in the runoff, in which case $c$ also does not win in this parallel universe), contradicting that $c$ is the unique winner. □

**Claim 3** *For each $j \leq t$, $Plu_{P_1^*}(s_j) \leq 3$.*

*Proof.* If this does not hold, then we let $j^*$ be an index that maximizes $Plu_{P_1^*}(s_{j^*})$. It follows that $Plu_{P_1^{2*}}(s_{j^*}) \geq 1$, because $Plu_{P_1^{1*}}(s_{j^*}) \leq 3$. However, by putting $s_{j^*}$ in the top position in a partial order in $P_1^2$, we are forcing $D(c, s_{j^*})$ to be reduced by 2, which means that $s_{j^*}$ defeats $c$ in their pairwise election. Moreover, because, by Claim 2, one of the $s_j$ must enter the runoff, and because $s_{j^*}$ has the maximum plurality score among the alternatives in $\mathcal{S}_V$, $s_{j^*}$ must be in the runoff in one of the parallel universes. However, $c$ cannot win in this parallel universe, which contradicts the assumption that $c$ is the unique winner. □

**Claim 4** $Plu_{P_1^*}(d) = 0$, $Plu_{P_1^*}(e) \leq 2q/3$.

*Proof.* It follows from Claim 3 that for each $j \leq t$, $Plu_{P_1^* \cup P_2}(s_j) \leq q + 3$. Therefore, by Claim 2 we must have that $Plu_{P_1^* \cup P_2}(d) \leq q + 2$ and $Plu_{P_1^* \cup P_2}(e) \leq q + 2$. The claim follows. □

**Claim 5** *For any $j \leq t$, if $Plu_{P_1^{2*}}(s_j) \geq 1$, then $Plu_{P_1^*}(s_j) \leq 2$.*

*Proof.* If $Plu_{P_1^{2*}}(s_j) \geq 1$ and $s_j$ enters the runoff in some parallel universe, then $c$ cannot win in that parallel universe. For the sake of contradiction, suppose $Plu_{P_1^*}(s_j) \geq 3$. By Claim 3 and Claim 2, $s_j$ enters the runoff in some parallel universe, which contradicts the assumption that $c$ is the unique winner. □

**Claim 6** *Let $X_1 = \{s_j : Plu_{P_1^{1*}}(s_j) > 0, Plu_{P_1^{2*}}(s_j) = 0\}$, and $X_2 = \{s_j : Plu_{P_1^{1*}}(s_j) = 0, Plu_{P_1^{2*}}(s_j) > 0\}$. We have $X_1 \cup X_2 = \mathcal{S}_V$ and $|X_1| = q/3$.*

*Proof.* Let $x_1 = |X_1|$, $x_2 = |X_2|$, and $x_3 = t - x_1 - x_2$. By Claim 5, for each $s_j \in \mathcal{S}_V \setminus (X_1 \cup X_2)$, $Plu_{P_1^{1*}}(s_j) = Plu_{P_1^{2*}}(s_j) = 1$. We recall that for each $O \in P_1^1$, the top-ranked alternative in any extension of $O$ must be either $d$ or an element in $\mathcal{S}_V$; for each $Q \in P_1^2$, the top-ranked alternative in any extension of $Q$ must be $d$, $e$, or an element in $\mathcal{S}_V$. We then use these observations to obtain two inequalities.

First, in order for $c$ to be the unique winner, $d$ cannot be in the top position in any vote in $P_1^{1*}$. Therefore, all of the $q$ top positions in $P_1^{1*}$ must be taken by alternatives in $\mathcal{S}_V$. Now, any alternative in $X_1$ can take at most three of these top positions; any alternative in $X_2$ takes none of these top positions by definition; and any alternative in $\mathcal{S}_V \setminus (X_1 \cup X_2)$ takes one of these top positions. It follows that $3x_1 + x_3 \geq q$.

Now, we apply a similar analysis to $P_1^{2*}$. In order for $c$ to be the unique winner, $e$ cannot be in the top position in more than $2q/3$ votes in $P_1^{2*}$, leaving at least $2t - 2q/3$ top positions to be filled. Now, any alternative in $X_1$ takes none of these top positions; any alternative in $X_2$ can take at most two of these top positions (Claim 5); and any alternative in $\mathcal{S}_V \setminus (X_1 \cup X_2)$ takes one of these top positions. It follows that $2x_2 + x_3 \geq 2t - 2q/3$.

By substituting $q$ in the second inequality by the $q$ in the first inequality, we obtain $2x_1 + 2x_2 + \frac{5}{3}x_3 \geq 2t$. We recall that $x_1 + x_2 + x_3 = t$. Therefore, $x_3 = 0$, $x_1 + x_2 = t$. Now the first inequality becomes $x_1 \geq q/3$ and the second inequality becomes $x_2 \geq t - q/3$. It follows from $x_1 + x_2 = t$ that $x_1 = q/3$ and $x_2 = t - q/3$. □

Based on all these claims, we can now construct a solution to the X3C instance. Let $X_1 = \{s_{j_1}, \ldots, s_{j_{q/3}}\}$. From Claim 3, Claim 6, $|P_1^{1*}| = q$, and the fact that every top position in $P_1^{1*}$ must be occupied by one of the alternatives in $X_1$, it follows that $S_{j_1}, \ldots, S_{j_{q/3}}$ is a solution to the X3C instance. Therefore, PW with respect to plurality with runoff is NP-complete. □

**Theorem 9** *NcW is coNP-complete with respect to plurality with runoff, even when the number of undetermined pairs in each vote is at most 4.*

*Proof.* We prove coNP-hardness by a reduction from X3C. Given an X3C instance $\mathcal{V} = \{v_1, \ldots, v_q\}$, $\mathcal{S} = \{S_1, \ldots, S_t\}$, we construct a NcW instance as follows.

**Alternatives:** $\{c, d\} \cup \mathcal{V} \cup E$, where $E = \{e_1, \ldots, e_{t(q+2)^3}\}$.

**First part $P_1$ of the profile:** for each $j \leq t$, we start with $O(d, S_j, c, \text{Others})$, and subsequently obtain a partial order $O_j$ by removing the orderings in $(\{d\} \cup S_j) \times \{c\}$.

**Second part $P_2$ of the profile:** $P_2 = P_2^1 \cup P_2^2$, where $P_2^1$ and $P_2^2$ are defined as follows.

- $P_2^1$: a set of $t(q+1) + q/3$ votes, such that $c$ is ranked in the top position $t + 1$ times; $d$ is ranked in the top position $q/3 - 1$ times; and for each $i \leq q$, $v_i$ is ranked in the top position $t$ times.

- $P_2^2$: we first obtain, according to Lemma 2, a profile $\hat{P}_2^2$ such that the pairwise score differences of $\{O(d, S_j, c, \text{Others}) : j \leq t\} \cup P_2^1 \cup \hat{P}_2^2$ satisfy the following conditions.

    1. $D(c, d) = 2t + 1$;

2. for all $i \leq q$, $D(v_i, c) = 3$.

By Lemma 2, the size of $\hat{P}_2^2$ is polynomial in $t + q$. Next, we obtain $P_2^2$ from $\hat{P}_2^2$ by raising an alternative in $E$ to the top position in each vote, in such a way that each vote in $P_2^2$ ranks a different alternative in the top position.

We recall that for any profile $P$ and any alternative $c'$, $Plu_P(c')$ denotes the number of times that $c'$ is ranked in the top position in $P$. We make the following observations on $\{O(d, S_j, c, \text{Others}) : j \leq t\} \cup P_2$.

- $D(c, d) = 2t + 1$, and for all $i \leq q$, $D(v_i, c) = 3$;

- $Plu(c) = t + 1$, $Plu(d) = t - 1 + q/3$; for each $i \leq q$, $Plu(v_i) = t$; for each $e \in E$, $Plu(e) \leq 1$.

It follows from the observations that in any extension of $P_1 \cup P_2$, $c$ must enter the runoff; also, in any extension, $c$ defeats $d$ in the pairwise election. Let $P_1^* \cup P_2$ (where $P_1^*$ is an extension of $P_1$) be a profile in which $c$ is not a co-winner. We must have that $d$ does not enter the runoff, which means that $Plu_{P_1^* \cup P_2}(d) \leq t - 1$. It follows that $c \succ d$ in at least $q/3$ votes in $P_1^*$. However, by ranking $c \succ d$ in a partial order $O_i$, we are forcing $c \succ S_i$. Now, the pairs of alternatives that enter the runoff (in parallel universes) are $(c, v_1), \ldots, (c, v_q)$. Since $c$ loses in all these pairwise elections in the runoff (because, by assumption, $c$ is not a co-winner), we must have that for each $v_j$, $c \succ v_j$ in at most one vote in $P_1^*$. Hence, a solution to the complement of the NcW instance naturally corresponds to a solution to the X3C instance. Conversely, a solution to the X3C instance corresponds to a solution to the complement of the NcW instance. Therefore, NcW with respect to plurality with runoff is coNP-complete. □

## 5. Polynomial-time Algorithms for Possible and Necessary Winner Problems

In this section we present polynomial-time algorithms for (1) NW and NcW with respect to all positional scoring rules, maximin, and Bucklin, (2) PcW and NW with respect to plurality with runoff. We recall that PW is NP-complete (Theorem 8) and NcW is coNP-complete (Theorem 9), both with respect to plurality with runoff.

We note that positional scoring rules, maximin, and Bucklin are all based on some type of scores, so if we can find an extension of the partial orders to linear orders so that the score of $c$, denoted by $S(c)$, is no more than the score of another alternative $w$, then $c$ is not the (unique) winner in this profile, and hence $c$ is not a necessary winner. Therefore, in the following algorithms for these rules, we check all alternatives $w \neq c$, and try to make $S(c) - S(w)$ as low as possible on a vote-by-vote basis (or equivalently, make $S(w) - S(c)$ as high as possible). For each vote $O$ (partial order), there can be two cases. In the first case, $c \not\succ_O w$. In this case, we only need to consider $c$ and $w$ separately, raising $w$ as high as possible and lowering $c$ as low as possible. (This part of the algorithm has already been considered in Konczak & Lang, 2005.) The following example, Example 3, illustrates this idea.

**Example 3** *A partial order $O$ is illustrated in Figure 4 (a). Let $c = c_2$ and $w = c_5$. Since $c_2 \not\succ_O c_5$, we can raise $c_5$ as high as possible while lowering $c_2$ as low as possible, as shown in Figure 4 (b).*

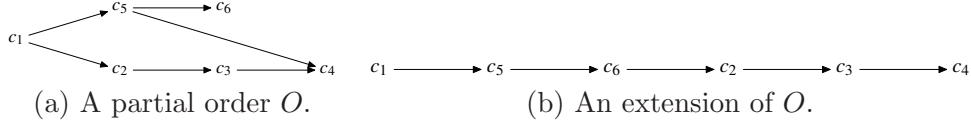

(a) A partial order $O$.    (b) An extension of $O$.

Figure 4: A partial order and its extension.

In the second case, $c \succ_O w$. This case is more complicated, and in what follows we show how to minimize $S(c) - S(w)$ for positional scoring rules, maximin, and Bucklin. For plurality with runoff, we convert PcW into a maximum flow problem to solve it; this also gives an algorithm for NW, simply by checking whether any other alternative is a possible co-winner (see Proposition 1).

In this section, the input consists of $\mathcal{C} = \{c, c_1, \ldots, c_{m-1}\}$, $c$ (the alternative for which we wish to decide whether or not it is a necessary (co-)winner), a profile $P_{poset}$ of $n$ partial orders over $\mathcal{C}$, and the voting rule $r$.

We first define some notation that will be used in the algorithms.

**Definition 8** *Given a partial order $O$ and an alternative $c$, let $Up_O(c) = \{c' \in \mathcal{C} : c' \succeq_O c\}$ and $Down_O(c) = \{c' \in \mathcal{C} : c \succeq_O c'\}$. Given another alternative $w$ such that $c \succ_O w$, let $O$'s $c \succ w$ block be defined as follows: $Block_O(c, w) = \{c' \in \mathcal{C} : c \succeq_O c' \succeq_O w\}$.*

That is, $Up_O(c)$ is the set of alternatives that are weakly preferred to $c$ in $O$ (including $c$ itself), and $Down_O(c)$ is the set of alternatives that $c$ is weakly preferred to in $O$ (including $c$ itself). If $c \succ_O w$, then $Block_O(c, w)$ is the set of all the alternatives, including $c$ and $w$, that are ranked between $c$ and $w$. It is easy to check that for any partial order $O$, and any pair of alternatives $c, w$ (with $c \succ_O w$), $Block_O(c, w) = Down_O(c) \cap Up_O(w)$.

**Example 4** *Let $O$ be the partial order illustrated in Figure 4 (a). We have that $Up_O(c_2) = \{c_1, c_2\}$, $Up_O(c_4) = \{c_1, c_2, c_3, c_4, c_5\}$, $Down_O(c_2) = \{c_2, c_3, c_4\}$, $Down_O(c_4) = \{c_4\}$, and $Block_O(c_2, c_4) = \{c_2, c_3, c_4\}$.*

The notion of a block is useful for the following reason. In the algorithm, we want to think about an extension of the partial orders in which $w$ does as well as possible, and $c$ does as poorly as possible. When $c \succ_O w$ in some partial order $O$, we cannot rank $c$ below $w$; but at least it makes sense to have as few alternatives between them as possible. The alternatives in the block are exactly the ones that need to be between them; we will rank the other alternatives outside of the block. Then, the question is where to position the block, and we will "slide" the block through the ranking.

Now we are ready to present the algorithms. We note that given a partial order $O$, computing the $Up_O$ and $Down_O$ sets takes polynomial time. Let $\vec{s}_m$ denote the scoring vector of the positional scoring rule.

**Algorithm 1 (Computing NW with respect to a positional scoring rule)**

1. For each partial order $O \in P_{poset}$ and each alternative $c$, compute $\text{Up}_O(c)$ and $\text{Down}_O(c)$.

2. Repeat Steps 3a–c for each $w \neq c$:

   **3a.** Let $S(w) = S(c) = 0$.

   **3b.** For each partial order $O$ in $P_{poset}$,
   - if $c \not\succ_O w$, then (following Example 3) the lowest possible position for $c$ is the $m + 1 - |\text{Down}_O(c)|$th position, and the highest possible position for $w$ is the $|\text{Up}_O(w)|$th position, so we add the scores $\vec{s}_m(|\text{Up}_O(w)|)$ and $\vec{s}_m(m + 1 - |\text{Down}_O(c)|)$ to $S(w)$ and $S(c)$, respectively;
   - if $c \succ_O w$, then the highest that we can slide $O$'s $c \succ w$ block (as measured by $c$'s position, which is at the top of the block) is position $|\text{Up}_O(w) \setminus \text{Down}_O(c)| + 1$ (if an alternative $a$ is ranked above $w$ in the partial order, then we will place it above $c$, unless the partial order ranks $c$ above $a$), and the lowest (as measured by $w$'s position, which is at the bottom of the block) is position $m - |\text{Down}_O(c) \setminus \text{Up}_O(w)|$ (if an alternative $a$ is ranked below $c$ in the partial order, then we will place it below $w$, unless the partial order ranks $a$ above $w$). Any position between these extremes is also possible. We find the position that minimizes the score of $c$ minus the score of $w$, then add the scores that $c$ and $w$ get for these positions to $S(c)$ and $S(w)$, respectively.

   **3c.** If the result is that $S(w) \geq S(c)$, then output that $c$ is not a necessary winner (terminating the algorithm).

3. Output that $c$ is a necessary winner (if we reach this point).

The algorithm for computing NcW is obtained simply by checking whether $S(w) > S(c)$ in Step 4.

**Proposition 3** *Algorithm 1 checks whether or not $c$ is a necessary winner for $P_{poset}$ with respect to a given positional scoring rule in polynomial time.*

*Proof.* It is equivalent to check whether there exists an extension $P$ of $P_{poset}$ and an alternative $w \neq c$, such that $s(P, w) \geq s(P, c)$—that is, whether $c$ is not a necessary (unique) winner. To this end, for each $O \in P_{poset}$, we maximize $s(V_O, w) - s(V_O, c)$ over all extensions $V_O$ of $O$.

We recall that for each $i \leq m$, $\vec{s}_m(i)$ is the score of the alternative that is ranked at the $i$th position. For any extension $V_O$ of $O$, $s(V_O, w) \leq \vec{s}_m(|\text{Up}_O(w)|)$ (because $w$ cannot be ranked higher than the $|\text{Up}_O(w)|$th position) and $s(V_O, c) \geq \vec{s}_m(m + 1 - |\text{Down}_O(c)|)$ (because $c$ cannot be ranked lower than the $(m + 1 - |\text{Down}_O(c)|)$th position). These two bounds can be achieved if $c \not\succ_O w$: for every $d \in \mathcal{C} \setminus \text{Up}_O(w)$, we add $w \succ d$ to $O$; and for every $d \in \mathcal{C} \setminus \text{Down}_O(c)$, we add $d \succ c$ to $O$. We obtain a partial order $O'$ this way, and we let $V_O$ be an (arbitrary) linear order that extends $O'$. It follows that $s(V_O, w) - s(V_O, c) = \vec{s}_m(|\text{Up}_O(w)|) - \vec{s}_m(m + 1 - |\text{Down}_O(c)|)$.

However, if $c \succ_O w$, there may not exist $V_O$ in which $s(V_O, w) = \vec{s}_m(|\text{Up}_O(w)|)$ and $s(V_O, c) = \vec{s}_m(m + 1 - |\text{Down}_O(c)|)$ hold simultaneously. We note that in any $V_O^*$ that

maximizes $s(V_O, w) - s(V_O, c)$, the only alternatives between $c$ and $w$ must be those in $\text{Block}_O(c, w)$. Therefore, for each $d \in \mathcal{C}$ such that $d \succ_O w$ and $c \not\succ_O d$, we must have that $d \succ_{V_O^*} c$; and for each $d \in \mathcal{C}$ such that $c \succ_O d$ and $d \not\succ_O w$, we must have that $w \succ_{V_O^*} d$. It follows that $s(V_O, w) - s(V_O, c) \leq \max_l(\vec{s}_m(l + |\text{Block}_O(c, w)| - 1) - \vec{s}_m(l))$, where $l$ ranges between $|\text{Up}_O(w) \setminus \text{Down}_O(c)| + 1$ and $m - |\text{Down}_O(c) \setminus \text{Up}_O(w)|$. Let $V'_O$ be an extension of $\mathcal{O}$ restricted to $\mathcal{C} \setminus \text{Block}_O(c, w)$ in which $\text{Up}_O(w) \setminus \text{Down}_O(c)$ is ranked at the top and $\text{Down}_O(c) \setminus \text{Up}_O(w)$ is ranked at the bottom. For each $d \in \mathcal{C} \setminus (\text{Up}_O(w) \cup \text{Down}_O(c))$ and each $d' \in \text{Block}_O(c, w)$, we must have $d \not\succ_O d'$ and $d' \not\succ_O d$. Therefore, for each $|\text{Up}_O(w) \setminus \text{Down}_O(c)| + 1 \leq l \leq m - |\text{Down}_O(c) \setminus \text{Up}_O(w)|$, we can put $\text{Block}_O(c, w)$ between the $(l-1)$th position and the $l$th position in $V'_O$, to obtain a linear order that extends $O$.

This proves the correctness of Step 3b, which computes $\max_{V_O}(s(V_O, w) - s(V_O, c))$. It follows that the algorithm correctly checks whether or not $c$ is a necessary winner. □

We now move on to the maximin rule. We note that $c$ is not a necessary winner for $P_{poset}$ with respect to maximin if and only if there exists a profile of linear orders $P$ extending $P_{poset}$, and two alternatives $w$ and $w'$, such that for all alternatives $d$, $N_P(w, d) \geq N_P(c, w')$. We recall that $N_P(w, d)$ is the number of votes in $P$ where $w \succ d$. Therefore, our algorithm considers all pairs $(w, w')$, and then checks whether there exists an extension of the input partial orders for which the inequality holds for all alternatives $d$. To perform such a check, in each partial order, we would like to rank $w'$ ahead of $c$, and also to rank $w$ as high as possible. However, these two objectives may conflict: it may be the case that if we rank $c$ ahead of $w'$, then we can rank $w$ higher than in the case where we rank $w'$ ahead of $c$. In this case, we first place $w'$ ahead of $c$, and then rank $w$ as high as possible under this additional constraint. This works for the following reason. Let $O \in P_{poset}$ be a partial order where $c \not\succ_O w'$ and $w' \not\succ_O c$; let $V$ be an arbitrary extension of $O$ in which $w' \succ_V c$ and let $V'$ be an arbitrary extension of $O$ in which $c \succ_{V'} w'$. For any $d \in \mathcal{C}$, we have that $N_{\{V\}}(w, d) - N_{\{V\}}(c, w') \geq 0 \geq N_{\{V'\}}(w, d) - N_{\{V'\}}(c, w')$, which means that enforcing $w' \succ c$ is always at least as good as enforcing $c \succ w'$.

**Algorithm 2 (Computing NW with respect to maximin)**

1. For each partial order $O \in P_{poset}$ and each alternative $c$, compute $\text{Up}_O(c)$.

2. Repeat 3a–c for all pairs $w, w'$, where $c \neq w$ and $c \neq w'$.

   **3a.** Let $S(c, w') = 0$, and for each alternative $d \neq w$, let $S(w, d) = 0$.

   **3b.** For each partial order $O$ in $P_{poset}$,
   - if $c \not\succ_O w'$, then add $w' \succ c$ to $O$ and raise $w$ as high as possible; for each $d \neq w$, if, in the resulting vote, $w$ is ahead of $d$ (that is, $d \notin \text{Up}_O(w)$ and if $c \in \text{Up}_O(w)$, then $d \notin \text{Up}_O(w')$), then add 1 to $S(w, d)$.
   - if $c \succ_O w'$, then raise $w$ as high as possible; add 1 to $S(c, w')$; for each $d \neq w$, if, in the resulting vote, $w$ is ahead of $d$ (that is, $d \notin \text{Up}_O(w)$), then add 1 to $S(w, d)$.

   **3c.** Check if for all $d \neq w$, $S(w, d) \geq S(c, w')$; if the answer is yes, then output that $c$ is not a necessary winner (terminating the algorithm).

4. Output that $c$ is a necessary winner.

The algorithm for computing NcW with respect to maximin is similar: the only modification is that in Step 3, we check if for all alternatives $d \neq w$, $S(w, d) > S(c, w')$.

**Proposition 4** *Algorithm 2 checks whether or not $c$ is a necessary winner for $P_{poset}$ with respect to maximin in polynomial time.*

*Proof.* The function $S(x, y)$ computed in the algorithm is the number of times $x$ is preferred to $y$ in an extension of $P_{poset}$. For any partial order $O$, we let $V_O$ be the extension computed in Step 3b. Let $g(V, d) = N_V(w, d) - N_V(c, w')$. We next prove that for each $d \neq w$ and each extension $V'_O$ of $O$, $g(V_O, d) \geq g(V'_O, d)$. If $c \not\succ_O w'$ and $c \succ_{V'_O} w'$, then $g(V'_O, d) \leq 0 \leq g(V_O, d)$ (because $N_{V_O}(c, w') = 0$ and $N_{V'_O}(c, w') = 1$). If $c \not\succ_O w'$ and $w' \succ_{V'_O} c$, then $N_{V'_O}(c, w') = N_{V_O}(c, w')$. We note that $V_O$ is obtained by raising $w$ as high as possible in $O$ while $w' \succ c$, which means that $N_{V'_O}(w, d) \leq N_{V_O}(w, d)$. It follows that $g(V_O, d) \geq g(V'_O, d)$. Similarly, if $c \succ_O w'$, then we also have that for all $d \neq w$, $N_{V'_O}(w, d) \leq N_{V_O}(w, d)$.

Therefore, for any extension $P$ of $P_{poset}$ and any $d \neq w$, $S(w, d) - S(c, w') = N_P(w, d) - N_P(c, w') \leq \sum_{O \in P_{poset}} g(V_O, d)$, and when $P$ is the profile computed in Step 3b, the inequality becomes an equality. It follows that the algorithm is correct. □

Now we move on to the Bucklin rule. We note that $c$ is not a necessary winner of $P_{poset}$ with respect to Bucklin, if and only if there exists an extension $P$ of $P_{poset}$ and an alternative $w$, such that either $w$'s Bucklin score is 1, or there exists $2 \leq k \leq m$, such that $w$ is among the top $k$ for more than $\frac{n}{2}$ votes (meaning that $w$'s Bucklin score is no more than $k$), and $c$ is among the top $k - 1$ for at most $\frac{n}{2}$ votes (meaning that $c$'s Bucklin score is no less than $k$). Therefore, like Algorithm 1, the algorithm for Bucklin considers each alternative $w$, computes the possible positions for the blocks $\text{Block}_O(c, w)$, and then checks for all $k$ from 1 to $m$ whether the above condition can be made to hold.

In the algorithm, if $c \not\succ_{O_j} w$, then $\text{High}(j)$ is the highest position that $w$ reaches in an extension of $O_j$, and $\text{Low}(j)$ is the lowest position that $c$ reaches in an extension of $O_j$. If $c \succ_{O_j} w$, then $\text{High}(j)$ is the highest position of $c$ given that $c$ and $w$ are ranked as close to each other as possible, $\text{Low}(j)$ is the lowest position of $c$ given that $c$ and $w$ are ranked as close to each other as possible, and $\text{Length}(j)$ is the size of $\text{Block}_{O_j}(c, w)$. For any $i \leq m$ and any $d \in \{c, w\}$, let $S(i, d)$ denote the minimum number of times that $d$ is ranked in the top $i$ positions, where the minimum is taken over all optimal extensions of $P_{poset}$ (we will elaborate on the meaning of optimality later). $U(k)$ is the number of partial orders for which we will have to compute where to put the block $\text{Block}_{O_j}(c, w)$ to make $c$ not a necessary unique winner. That is, $U(k)$ is the number of partial orders for which there exists an extension in which $c$ is in the top $k - 1$ positions and $w$ is in the top $k$ positions, as well as another extension in which $c$ is not in the top $k - 1$ positions and $w$ is not in the top $k$ positions.

**Algorithm 3 (Computing NW with respect to Bucklin)**

1. For each partial order $O \in P_{poset}$ and each alternative $c$, compute $\text{Up}_O(c)$ and $\text{Down}_O(c)$.

2. Repeat Steps 3a–d for all $w \neq c$:

    **3a.** For each $j \leq n$, let $\text{High}(j) = \text{Low}(j) = \text{Length}(j) = 0$. For each $i \leq m$, let $S(i, c) = S(i, w) = U(i) = 0$.

**3b.** For each partial order $O_j$ in $P_{poset}$,
- if $c \not\succ_{O_j} w$, then let $\text{Length}(j) = 0$, and let $\text{High}(j) = |\text{Up}_{O_j}(w)|$, $\text{Low}(j) = m + 1 - |\text{Down}_{O_j}(c)|$;
- if $c \succ_{O_j} w$, then let $\text{Length}(j) = |\text{Block}_{O_j}(c,w)|$, $\text{High}(j) = |\text{Up}_{O_j}(w) \setminus \text{Down}_{O_j}(c)| + 1$, $\text{Low}(j) = m + 1 - |\text{Down}_{O_j}(c)|$.

**3c.** For each $k \leq m$, each $j \leq n$,
- if $\text{Length}(j) = 0$, then add 1 to $S(k,w)$ if $\text{High}(j) \leq k$, and add 1 to $S(k-1,c)$ if $\text{Low}(j) \leq k-1$;
- if $\text{Length}(j) > 0$, then add 1 to $S(k,w)$ if either $\text{Low}(j) + \text{Length}(j) - 1 \leq k$, or the following two conditions both hold: $\text{Low}(j) \leq k-1$ and $\text{High}(j) + \text{Length}(j) - 1 \leq k$. Also, add 1 to $S(k-1,c)$ if $\text{Low}(j) \leq k-1$; add 1 to $U(k)$ if $\text{Low}(j) > k-1$ and $\text{High}(j) + \text{Length}(j) - 1 \leq k$.

**3d.** If $S(1,w) + U(1) > \frac{n}{2}$, or there exists $2 \leq k \leq m$ such that $S(k,w) > S(k-1,c)$, $S(k-1,c) \leq \frac{n}{2}$, and $S(k,w) + U(k) > \frac{n}{2}$, then output that $c$ is not a necessary winner (terminating the algorithm).

4. Output that $c$ is a necessary winner.

The algorithm for computing NcW is obtained by making following changes to Steps 3c and 3d as follows.

**3c'.** For each $k \leq m$, each $j \leq n$,

- if $\text{Length}(j) = 0$, then add 1 to $S(k,w)$ if $\text{High}(j) \leq k$, and add 1 to $S(k,c)$ if $\text{Low}(j) \leq k$;

- if $\text{Length}(j) > 0$, then add 1 to $S(k,w)$ if either $\text{Low}(j) + \text{Length}(j) - 1 \leq k$, or the following two conditions both hold: $\text{Low}(j) \leq k$ and $\text{High}(j) + \text{Length}(j) - 1 \leq k$. Also, add 1 to $S(k,c)$ if $\text{Low}(j) \leq k$; add 1 to $U(k)$ if $\text{Low}(j) \geq k+1$ and $\text{High}(j) + \text{Length}(j) - 1 \leq k$.

**3d'.** If there exists $0 \leq l \leq U(1)$ such that $S(1,w) + l > \frac{n}{2} \geq S(1,c) + l$, or there exists $2 \leq k \leq m$ and $l \leq U(k)$ such that $S(k,w) + l > \frac{n}{2} \geq S(k,c) + l$, then output that $c$ is not a necessary co-winner (terminating the algorithm).

**Proposition 5** *Algorithm 3 checks whether or not $c$ is a necessary winner for $P_{poset}$ with respect to Bucklin in polynomial time.*

*Proof.* Similarly as in the case of positional scoring rules, for Bucklin, if $c \not\succ_O w$, then we can simply rank $c$ as low as possible while rank $w$ as high as possible, independently. On the other hand, if $c \succ_O w$, then we can without loss of generality place as few alternatives between $c$ and $w$ as possible, but the question is where to place the $c \succ w$ block. The algorithm will consider a particular $k$, and try to make it so that $w$ is among the top $k$ for more than half the votes, and $c$ is among the top $k - 1$ for at most half the votes. For a particular vote with $c \succ_O w$, depending on where the block is placed, either (1) $c$ is among the top $k - 1$ and $w$ is among the top $k$; or, (2) $c$ is among the top $k - 1$ and $w$ is not among the top $k$; or, (3) $c$ is not among the top $k - 1$ and $w$ is not among the top $k$. However, not all three of these possibilities may exist for a particular vote. The algorithm will never choose (2) unless that is the only option, so that the only difficult case is when a decision must be made between (1) and (3).

We recall that for any $i \leq m$ and any $d \in \{c, w\}$, $S(i, d)$ is the minimum number of times that $d$ is ranked within top $i$ positions, where the minimum is taken over all extensions of $P_{poset}$ that are consistent with the observations in the previous paragraph (specifically, option (2) is never chosen unless there is no other choice). $U(k)$ is the number of partial orders for which there exists an extension in which $c$ is ranked within top $k-1$ positions and $w$ is ranked within top $k$ positions, as well as an extension in which $c$ is not ranked within top $k-1$ positions and $w$ is not ranked within top $k$ positions (that is, we have a choice between (1) and (3)).

For each $k \leq m$, and each $j \leq n$, we consider how to extend $O_j$.

- If $c \not\succ_{O_j} w$, then the positions of $c$ and $w$ are already determined by our previous observations ($w$ is ranked as high as possible and $c$ is ranked as low as possible).

- If $c \succ_{O_j} w$ and $\text{High}(j) \geq k$, then $c$ cannot be ranked within top $k-1$ positions and $w$ cannot be ranked within top $k$ positions; therefore, we add 0 to $S(k-1, c)$ and $S(k, w)$.

- If $c \succ_{O_j} w$, $\text{High}(j) < k$ and $\text{High}(j) + \text{Length}(j) - 1 > k$, then $c$ can be ranked within top $k-1$ positions, but $w$ cannot be ranked within top $k$ positions. There are two sub-cases: (1) if $\text{Low}(j) \geq k$, then we rank $c$ in the $\text{Low}(j)$th position, and henceforth add 0 to both $S(k-1, c)$ and $S(k, w)$; (2) if $\text{Low}(j) < k$, then $c$ is inevitably ranked within top $k-1$ positions, and $w$ cannot be ranked within top $k$ positions, which means that we add 1 to $S(k-1, c)$ and 0 to $S(k, w)$.

- The final case is where $c \succ_{O_j} w$, $\text{High}(j) < k$ and $\text{High}(j) + \text{Length}(j) - 1 \leq k$. Again, there are two subcases: (1) if $\text{Low}(j) < k$, then it means that $c$ must be ranked within top $k-1$ positions. Therefore we rank $w$ in the top $k$ positions, and add 1 to both $S(k-1, c)$ and $S(k, w)$; (2) if $\text{Low}(j) \geq k$, then it means that we have three options for an extension of $O_j$, corresponding to the cases (1), (2), (3) discussed in the beginning of the proof.

  (1) $c$'s position is within top $k-1$ and $w$'s position is within top $k$.
  
  (2) $c$'s position is within top $k-1$ and $w$'s position is not within top $k$ (which implies that $\text{Length}(i) > 2$).
  
  (3) $c$'s position is not within top $k-1$ and $w$'s position is not within top $k$.

  As we already discussed, option (2) is suboptimal. Therefore, we add 0 to both $S(k-1, c)$ and $S(k, w)$, and add 1 to $U(k)$.

The only remaining decision is for how many of the votes corresponding to the number $U(k)$ to choose option (1) (as opposed to option (3)). This corresponds to Step 3d of the algorithm, where it checks whether there exists a way of choosing the number of extensions (but no more than $U(k)$) that choose (1) in such a way that $c$ is not the winner.

Therefore, the algorithm is correct. □

Finally, we consider the possible co-winner problem with respect to plurality with runoff. We will show that this problem can be solved in polynomial time. From this, it also follows that the necessary (unique) winner problem can be solved in polynomial time (Proposition 1). In contrast, we have already shown that for plurality with runoff, the possible unique winner problem is NP-complete (Theorem 8) and the necessary co-winner problem is coNP-complete (Theorem 9).

Our algorithm for determining whether $c$ is a possible co-winner is based on the following key observation: $c$ is a possible co-winner for $P_{poset}$ with respect to plurality with runoff if and only if there exists an extension of $P_{poset}$, denoted by $P^*$, an alternative $d \neq c$, and two natural numbers $l_1, l_2$, such that (1) $c$ is preferred to $d$ in at least half of votes (linear orders) in $P^*$, and (2) $Plu_{P^*}(c) = l_1$, $Plu_{P^*}(d) = l_2$, and for each alternative $c'$ ($c' \neq c$ and $c' \neq d$), $Plu_{P^*}(c') \leq \min\{l_1, l_2\}$. That is, $c$ and $d$ can enter the runoff (there could be other pairs of alternatives who enter the runoff in some parallel universe) and $c$ can then defeat $d$ in the runoff.

For each $i^* \leq m-1$, we let $\alpha_{i^*}$ denote the number of partial orders $O \in P_{poset}$ such that $c_{i^*} \succ_O c$. We recall that $Top(O)$ denote the set of alternatives $c'$ for which there exists at least one extension of $O$ where $c'$ is in the top position. Based on the observations in the previous paragraph, we will consider all possibilities for $l_1$, $l_2$, and $d$ (we will use $i^*$ to denote possibilities for the index of $d$), and solve a maximum flow problem instance for each possibility.[9] Specifically, for every $l_1, l_2 \leq n$ and every $i^* \leq m - 1$ (with $\alpha_{i^*} \leq n/2$), we define a maximum flow problem $F_{l_1, l_2, i^*}$ as follows (illustrated in Figure 5, in which $i^* = 1$).

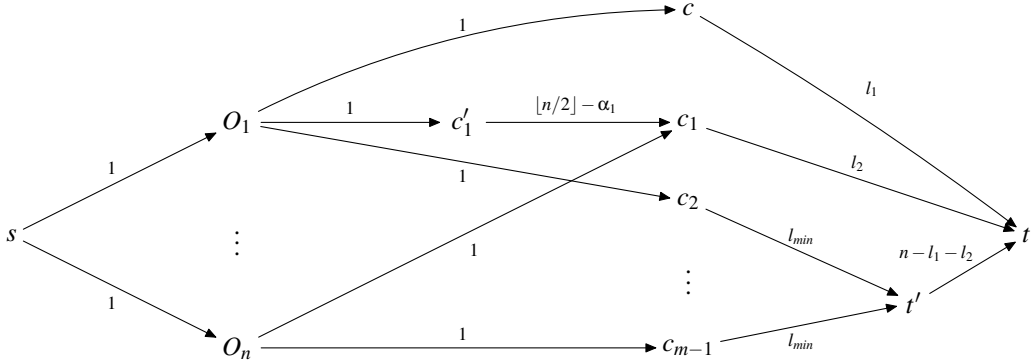

Figure 5: The maximum flow problem $F_{l_1, l_2, 1}$.

**Vertices:** $s$, $O_1, \ldots, O_n$, $c'_{i^*}$, $c, c_1, \ldots, c_{m-1}$, $t'$, $t$.

**Edges**: we have the following five types of edges.

- **Edges from $s$ to $\{O_1, \ldots, O_n\}$**: for every $i \leq n$, there is an edge $(s, O_i)$ with capacity 1.

---

9. Our original proof used a minimum cost flow problem, but one of the anonymous reviewers pointed out how to modify this approach into the simpler maximum flow approach presented here, as well as two papers (Gusfield & Martel, 2002; Russell & Walsh, 2009) where maximum flow problems were used to solve other election problems, for which we thank the reviewer.

– **Edges from $\{O_1, \ldots, O_n\}$ to $\{c'_{i^*}, c, c_1, \ldots, c_{m-1}\}$**: we have
  * for every $j \leq n$ and every $d \in \mathcal{C}$ such that $d \neq c_{i^*}$, if $d \in Top(O_j)$, then there is an edge $(O_j, d)$ with capacity 1;
  * for every $j \leq n$, if $c_{i^*} \in Top(O)$ and $c_{i^*} \succ_{O_j} c$, then there is an edge $(O_j, c_{i^*})$ with capacity 1;
  * for any $j \leq n$, there is an edge $(O_j, c'_{i^*})$ with capacity 1 if $c_{i^*} \in Top(O)$ and $c_{i^*} \not\succ_{O_j} c$.
– **Edge from $c'_{i^*}$ to $c_{i^*}$**: there is an edge $(c'_{i^*}, c_{i^*})$ with capacity $\lfloor n/2 \rfloor - \alpha_{i^*}$.
– **Edges from $\mathcal{C} \setminus \{c, c_{i^*}\}$ to $t'$**: for every $c' \in \mathcal{C} \setminus \{c, c_{i^*}\}$, we have an edge $(c', t')$ with capacity $l_{min} = \min\{l_1, l_2\}$.
– **Edges from $\{c, c_{i^*}, t'\}$ to $t$**: we have
  * an edge $(c, t)$ with capacity $l_1$;
  * an edge $(c_{i^*}, t)$ with capacity $l_2$;
  * an edge $(t', t)$ with capacity $n - l_1 - l_2$.

Next, we prove that $c$ is a possible co-winner for $P_{poset}$ with respect to plurality with runoff if and only if there exist $l_1, l_2 \leq n$ and $i^* \leq m-1$ such that $F_{l_1, l_2, i^*}$ has a solution in which the value of the flow is $n$.

Because all parameters in $F_{l_1, l_2, i^*}$ are integers, if there exists a solution to $F_{l_1, l_2, i^*}$, then there must also exists an integer solution. First, we show how to convert an integer solution to $F_{l_1, l_2, i^*}$ to a solution to the PcW problem with respect to plurality with runoff. Let $f$ be an integer solution to $F_{l_1, l_2, i^*}$, that is, $f$ : Vertices × Vertices → $\mathbb{Z}$. We construct an extension $P^* = (V_1^*, \ldots, V_n^*)$ of $P_{poset}$ as follows:

- for each $j \leq n$, if $f(O_j, c'_{i^*}) = 1$ then we let $V_j^*$ be an extension of $O_j$ in which $c_{i^*}$ is ranked in the top position;

- for each $j \leq n$ and each $d \in \mathcal{C} \setminus \{c_{i^*}\}$, if $f(O_j, d) = 1$ then we let $O_j^*$ be an extension of $O_j$ in which $d$ is ranked in the top position, and $c$ is ranked as high as possible.

Because the value of $f$ is $n$, the plurality score of $c$ is $l_1$ and the plurality score of $c_{i^*}$ is $l_2$, while the plurality score of $c_i$ ($i \neq i^*$) is at most $l_{min}$. Therefore, $c$ and $c_{i^*}$ enter the runoff together in one parallel universe. Now, the capacity constraint on the edge $(c'_{i^*}, c_{i^*})$ ensures that $c$ will win the runoff: the reason is that if we rank $c_{i^*}$ first in a vote in which we could have ranked $c$ ahead of $c_{i^*}$, then it will contribute 1 to the flow on this edge. Moreover, the capacity of the edge $(c'_{i^*}, c_{i^*})$ is $\lfloor n/2 \rfloor - \alpha_{i^*}$, which means that $c_{i^*} \succ c$ in at most $\alpha_{i^*} + (\lfloor n/2 \rfloor - \alpha_{i^*}) \leq n/2$ votes of $P^*$. Hence, $c$ is a co-winner for $P^*$.

Conversely, if there exists an extension $P^*$ of $P$ such that $c$ is a co-winner of $P^*$, then there exists a $c_{i^*}$ such that in some parallel universe, $\{c, c_{i^*}\}$ enter the runoff, and $c$ wins this runoff. Let $l_1, l_2$ be the plurality scores of $c, c_{i^*}$, respectively. Then, this extension can be converted to a solution to $F_{l_1, l_2, i^*}$ (we omit the details because they are similar to the details for the other direction).

Therefore, the following algorithm solves PcW with respect to plurality with runoff.

**Algorithm 4 (Computing PcW with respect to plurality with runoff)**

1. For each $O \in P_{poset}$, compute $Top(O)$ and $\text{Up}_O(c)$. For each $i \leq m-1$, let $\alpha_i = |\{O \in P_{poset} : c_i \in \text{Up}_O(c)\}|$.

2. Repeat Steps 3a–b for all $i \leq m-1$ and $l_1, l_2 \leq n$:

    **3a.** Construct the maximum flow problem $F_{l_1,l_2,i}$.

    **3b.** Solve $F_{l_1,l_2,i}$ by the Ford–Fulkerson algorithm (Cormen, Leiserson, Rivest, & Stein, 2001). If the maximum flow is $n$, then output that $c$ is a possible co-winner. Terminate the algorithm.

3. Output that $c$ is not a possible co-winner.

**Proposition 6** *Algorithm 4 checks whether or not $c$ is a possible co-winner for $P_{poset}$ with respect to plurality with runoff in polynomial time.*

We recall from the proof of Proposition 1 that $c$ is a necessary unique winner if and only if no other alternative is a possible co-winner. Therefore, we naturally obtain an algorithm for NW, simply by using Algorithm 4 to check if any alternative other than $c$ is a possible co-winner.

**Proposition 7** *Algorithm 4 can be used to check whether or not $c$ is a necessary unique winner for $P_{poset}$ with respect to plurality with runoff in polynomial time.*

## 6. Conclusion and Future Work

We considered the following problem: given a set of alternatives, a voting rule, and a set of partial orders, which alternatives are possible/necessary winners? That is, which alternatives would win for some/all extension of the partial orders? We considered the case where the votes are not weighted and the number of alternatives is not bounded. Table 1 in the introduction summarizes our results. These results hold whether or not the alternative must be the unique winner, or merely a co-winner, unless specifically mentioned.

In this paper, there was no restriction on the partial orders. However, if the reason that we have partial orders is that preferences are submitted as CP-nets, this introduces additional structure on the partial orders; that is, not all partial orders correspond to a CP-net. Hence, while our positive results would still apply, it is not immediately obvious that our negative results would still apply.

Another approach is to approximate the sets of possible/necessary winners. More precisely, we are asked to output a superset (respectively, subset) of possible (respectively, necessary) winners such that the size of the output set should be within a fixed ratio of the number of the possible (respectively, necessary) winners. Pini et al. (2007) proved the inapproximability of the set of possible/necessary winners for the single transferable vote rule (STV) rule. We conjecture that similar inapproximability results hold for most of the common voting rules studied in this paper (for which the possible/necessary winner problems are (co-)NP-complete).

## Acknowledgments

We thank Nadja Betzler, Jérôme Lang, Toby Walsh, the anonymous reviewers for AAAI-08 and JAIR, and all participants of the Dagstuhl Seminar 07431: *Computational Issues in Social Choice* for helpful discussions and comments. Lirong Xia is supported by a James B. Duke Fellowship and Vincent Conitzer is supported by an Alfred P. Sloan Research Fellowship. This work is supported by NSF under award numbers IIS-0812113 and CAREER 0953756.